\documentclass{aa}  
\usepackage{graphicx}
\usepackage{txfonts}
\usepackage{natbib}
\usepackage[figuresright]{rotating}
\usepackage{aalongtable,lscape}

\def\ergs{\ensuremath{{\rm erg\,s^{-1}}}}                   

\def\lbol{\ensuremath{\mathit{L}_{\rm bol}}}              
\def\mdot{\ensuremath{\dot{M}}}                          
\def\etacar{$\eta$\,Carinae\,\,}

\def\tr{{Trumpler\,16\,}}
\def\chandra{{\em Chandra\,}}

\def\xmm{{\em XMM-Newton\,}}

\def\Lx{{L$_{\rm x}$}\,}
\def\Lbol{{L$_{\rm bol}$}\,}

\def\Av{A$_{\rm v}$\,\,}

\def\Av{\mbox{A$_{\rm v}$\,}}
\begin{document}
   \title{An X-ray survey of low-mass stars in Trumpler\,16 with Chandra}
   
   \author{J.F. Albacete-Colombo\inst{1,2}
     \and
     F. Damiani\inst{2}	
     \and
     G. Micela\inst{2}
     \and
     S. Sciortino\inst{2}
     \and
     F.\,R. Harnden, Jr.\inst{3}
   }
   
   \offprints{JFAC - email: albacete@fcaglp.unlp.edu.ar}

   \institute{1- Centro Universitario Regional Zona Atl\'antica -
   Univ. COMAHUE, Monse\~nor Esandi y Ayacucho (CP\,8500), Viedma,
   Argentina.\\ 2- INAF-Osservatorio Astronomico di Palermo, Piazza
   del Parlamento 1, I-90134, Palermo, Italy.\\ 3- AF Smithsonian
   Astrophysical Observatory, 60 Garden St., Cambridge, MA 02138,
   USA.\\\email{albacete@fcaglp.unlp.edu.ar} }

   \date{Received -----; accepted -----}
   
    \abstract{}
   {We identify and characterize low-mass stars in the $\sim$\,3\,Myr
   old \tr region by means of a deep \chandra X-ray observation, and
   study their optical and near-IR properties. We compare X-ray
   activity of \tr stars with known characteristics of Orion and
   Cygnus\,OB2 stars.}{We analyzed a 88.4 ksec \chandra\,ACIS-I
   observation pointed at the center of \tr. Because of diffuse X-ray
   emission, source detection was performed using the PWDetect code
   for two different energy ranges: 0.5-8.0\,keV and 0.9-8.0\,keV.
   Results were merged into a single final list. We positionally
   correlate X-ray sources with optical and 2MASS catalogues. Source
   events were extracted with the IDL-based routine ACIS-Extract.
   X-ray variability was characterized using the Kolmogorov-Smirnov
   test and spectra were fitted by using XSPEC. X-ray spectra of
   early-type, massive stars were analyzed individually.}{Our list of
   X-ray sources consists of 1035 entries, 660 of which have near-IR
   counterparts and are probably associated with \tr members. From
   near-IR color-color and color-magnitudes diagrams we compute
   individual masses of stars and their \Av values. The cluster median
   extinction is \Av=3.6 mag, while OB-type stars appear less
   absorbed, having \Av=2.0 mag. About 15\% of the near-IR
   counterparts show disk-induced excesses. X-ray variability is found
   in 77 sources, and typical X-ray spectral parameters are N$_{\rm
   H}\sim$5.37$\times$10$^{21}$ cm$^{-2}$ and kT$\sim$1.95 keV, with
   1$\sigma$ dispersions of 0.45 dex and 0.8 keV, respectively. OB
   stars appear, softer with a median kT$\sim$0.65 keV. The median
   X-ray luminosity is 6.3$\times$10$^{30}$ \ergs\,, while variable
   sources show a larger median \Lx value of 13$\times$10$^{30}$
   \ergs. OB-stars have an even higher median \Lx of
   80$\times$10$^{30}$ \ergs, about 10 times that of the low-mass
   stars.}{The \tr region has a very rich population of low-mass X-ray
   emitting stars. An important fraction of its circumstellar disks
   survive the intense radiation field of its massive stars. Stars
   with masses 1.5-2.5 M$_\odot$ display X-ray activity similar to
   that of stars in Cyg\,OB2 but much less intense than observed for
   Orion Nebula Cluster members.}


\maketitle

\section{Introduction} 
\label{intro}

The Carina nebula region (NGC\,3372) is one of the most massive star
formation regions of the Galaxy. It is associated with a giant H{\sc
ii} region spanning about 4 deg$^2$ of the sky and being bisected by a
prominent V-shaped dark gas and dusty lane. This prominent young
structure is not as compact as some of the other young galactic 
clusters, but seemingly related to a spiral feature. In this
direction, we are looking almost tangentially to the now recognized
Carina-Sagittarius spiral arm, at the edge of a giant molecular cloud
extending over about 130 pc and with a content in excess of
5$\times$10$^5$ solar masses \citep{1988ApJ...331..181G}. The
concentration of massive stars (i.e. M$\geq$20 M$_\odot$) interacts
with the parent giant molecular cloud of the region, leading to
triggered star formation events on intermediate to lower masses
\cite[e.g.][]{2004MNRAS.351.1457S}.\\ This region harbors several open
clusters and/or star concentrations (Trumpler 14, 15 and 16; Collinder
228 and 232; Bochum 10 and 11) containing more than 60 known O-type
stars \citep{1995RMxAC...2...57F}. Large cavities within the giant
molecular cloud are supposed to be carved out by the Tr\,14 and 16
open clusters, which contain most of massive stars of the region. In
particular Tr\,16 includes three rare main-sequence O3 stars, the
Wolf-Rayet (WR) star HD93162 and the famous luminous blue variable
(LBV) \etacar. There is a historical controversy about the distance
and age of Tr\,14 and Tr\,16 \citep{1995RMxAC...2...51W}. For
instance, from extensive spectroscopy and photometry
\cite{1993AJ....105..980M} find 3.2 kpc for both clusters. However,
photometric studies are strongly affected by differential extinction
in the region and peculiar reddening, and so the derived distance  are
different. An example is the \cite{2004A&A...418..525C} work, who for
different R=\Av/E(B-V) values (3.48 and 4.16 for Tr\,16 and Tr\,14
regions), compute distances of 4.0 kpc and 2.5 kpc, respectively. A
more reliable distance (2250$\pm$180 pc), was derived from proper
motion and Doppler velocities of the expanding \etacar Homunculus
using HST-STIS\footnote{Data from Hubble Space Telescope (HST) with
the Space Telescope Imaging Spectrograph (STIS)} observations
\citep{1997ARA&A..35....1D}. Recent work \citep{2003MNRAS.339...44T}
derives a common distance DM=12.14 (2.7 kpc) and an age between
$\sim$1\,Myr and 3\,Myr, for Tr14 and Tr16, respectively. For this
study, we adopt for \tr a distance of 2250 pc and an age of 3\,Myr.
This young age agrees with the \cite{2000ApJ...532L.145S} results, who
report the existence of several embedded IR sources where star
formation might be active. Also, \cite{2001ApJ...549..578D} confirm
clear evidence of pre-main sequence (PMS) stars in the region, while
\cite{2001MNRAS.327...46B} have identified two compact H{\sc ii}
regions possibly linked to very young O-type stars. Finally,
\cite{2004MNRAS.355.1237H} report a compact cluster of infrared
PMS-stars in Tr\,16.

Of the existing methods to identify young stellar populations, the use
of X-ray emission is perhaps the least biased
\citep{2002ApJ...574..258F}. While in main-sequence (MS) stars, from
late A to M dwarfs, X-rays are believed to originate from the hot
coronal gas that is heated by stellar dynamo magnetic fields
\citep{1987ApJ...315..687M}, for late type Pre-MS stars (T-Tauri stars
(TTSs)) X-ray emission is attributed to solar-like coronal activity
but elevated by a factor of 10$^3$-10$^4$ \citep{1999ARA&A..37..363F}.
Several authors suggested the possibility of detecting early pre-main
sequence (PMS) objects through their hard X-ray emission escaping the
highly obscured regions (see \cite{1992AJ....104..758W,
1997PASJ...49..461K, 1997ApJ...486L..39H, 2002ApJ...579L..95H}).
Recently, X--ray surveys have been successful in identifying the young
and pre-MS population in star-forming regions, including: $i-$ deeply
embedded Class I young stellar objects (YSOs), $ii-$ low-mass T-Tauri
PMS stars, $iii-$ intermediate-mass Herbig Ae/Be PMS stars, $iv-$
zero-age MS stars. Moreover, X-ray emission from low-mass pre-MS stars
usually exhibits a strong variability that helps to confirm
membership.

On last decade, X-ray observations of young stars on star-forming
regions were intensified thanks to the high spatial resolution and the
improved broad-band ([0.2-12.0] and [0.5-10.0] keV) effective area of
the \xmm\, and \chandra\, satellites. A first X-ray survey in the
Carina region by \cite{2003MNRAS.346..704C} was performed on the basis
of two early \xmm\, observations (rev\,\#115 and \#116) centered on
\etacar. Because of the spatial resolution of the
EPIC\footnote{European Photon Image Camera has about six times less
spatial resolution than \chandra ACIS-I camera.} camera and relatively
short exposure time of the observations ($\sim$35 ksec), they detected
only 80 X-ray sources, most of them related to the massive OB-type
stars with \Lx$\sim$10$^{32}$-10$^{34}$ \ergs. Before the observation
used here, three \chandra\, observations were obtained on this region,
two (obsId\,50 and 1249) in the timed exposure mode, and the third
(obsId\,51) in the continuous clocking mode, which produces no image.
Using only observation obsId\,1249, \cite{2003ApJ...589..509E}
presented luminosities and hardness ratios of the hot stars in Tr\,16,
and part of Tr\,14. Low-resolution X-ray spectra of luminous sources
were discussed by \cite{2004ApJ...612.1065E}. However the short
exposure time of such observation ($\sim$ 9.5 ksec) was a serious
limitation for the study of intermediate- and low mass stellar
population of the region. This limitation exists even if obsId. 50 and
obsId. 1249 are combined \citep{2007ApJ...656..462S}, reaching
completeness just at X-ray luminosity (\Lx) of
$\sim$7$\times$10$^{31}$ \ergs, i.e. the X-ray  emission level typical
of single O- and early B-type stars.

In this paper we present results of the analysis of the deepest X-ray
observation ever done in this region ($\sim$90 ksec). Section
\ref{obs} gives details on the observation and data reduction
procedures. Section\,3 explains the method used to detect the sources,
photon extraction and the construction of the catalog. In section\,4
we present results of the cross-correlation with existing near-IR and
optical catalogs of objects and their characterization based on their
color-color (CC) and color-magnitude (CM) diagrams. Section\,5
presents a statistical study of variability in the X-ray domain.
Section\,6  is dealing with results of the analysis of extracted X-ray
spectra.  In section\,7 we discuss X-ray luminosities of stars and
compare them statistically with the X-ray source population of ONC and
Cygnus\,OB2 star forming regions. In section\,8 we discuss X-ray and
stellar parameter of O- and early B- type stars. Finally, in
section\,9 we give a summary and draw conclusions of the paper.

\section{X-ray observations}
\label{obs}

\begin{figure*}[ht] \centering
\includegraphics[width=8.6cm,angle=0]{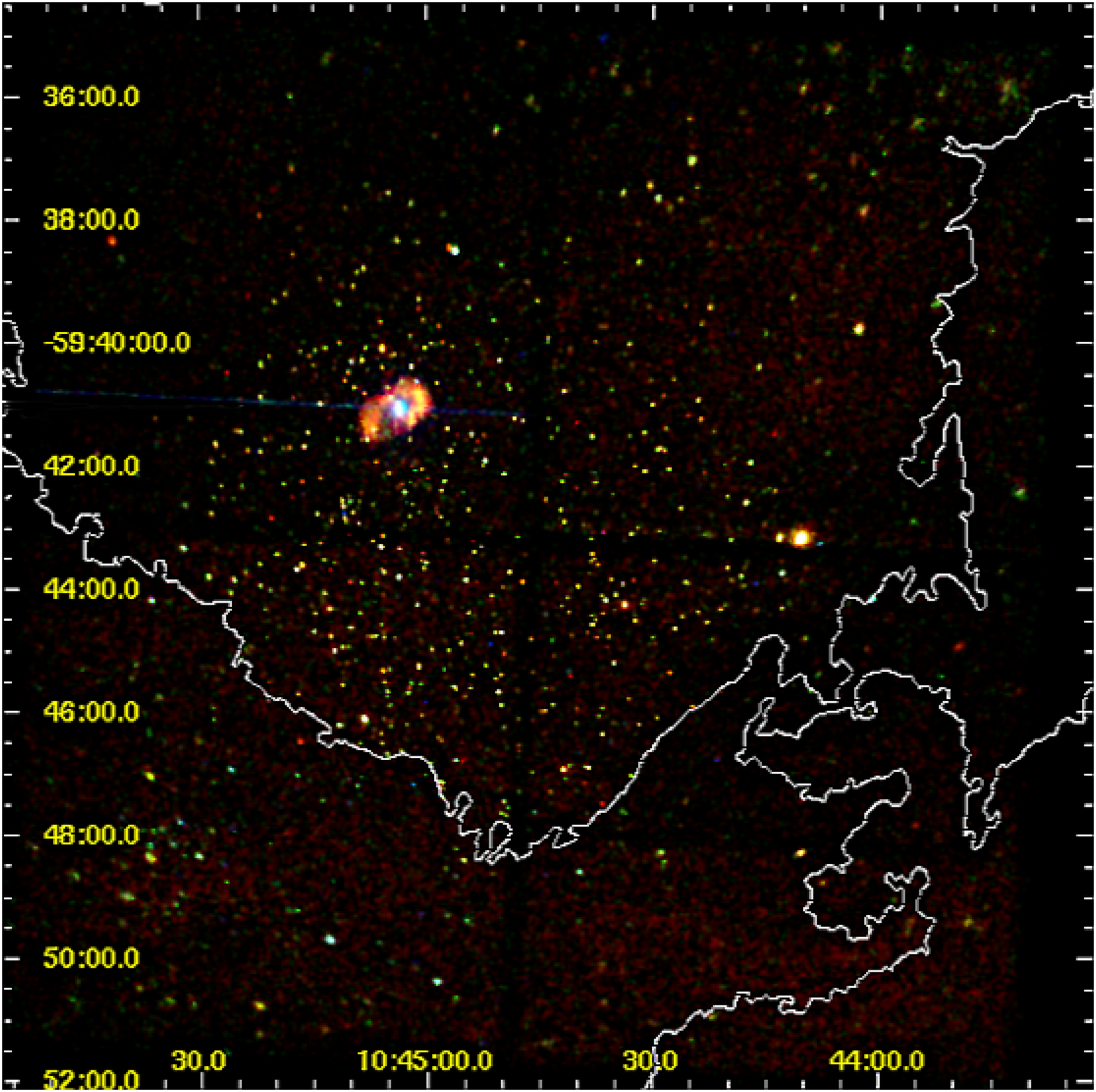}
\includegraphics[width=8.5cm,angle=0]{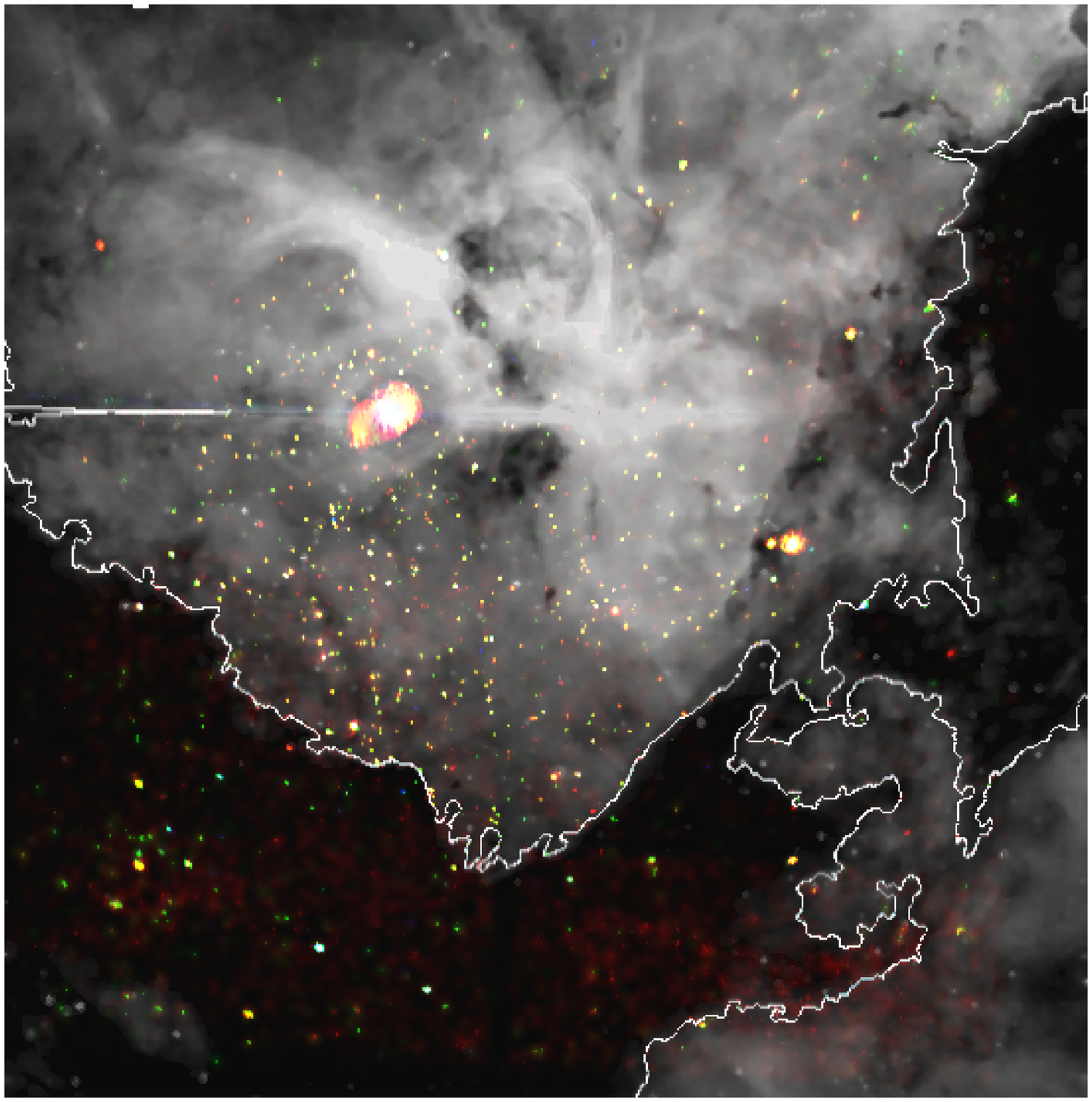} \caption{Left:
Color-coded ACIS-I image of the 17'\,$\times$\,17' field in \tr (see
color version in the electronic  edition). Kernel smoothing was
applied to highlight point sources. Energy bands for the RGB image are
[0.5:1.5], [1.5:2.2], and [2.2:8.0] keV for the red, green, and blue
colors, respectively. Contour lines show the spatial distribution of
the dark V-shaped dust lane. Right: ACIS-I image composed with a
H$_\alpha$ image. This image clearly show the sharp cut, along the
V-shape structure, in the spatial distribution of X-ray sources.}
\label{acis} \end{figure*}

\tr was observed with the ACIS detector on board the {\it Chandra
X-ray Observatory} (CXO) \citep{2002PASP..114....1W} on 2006 August
31\footnote{Observation start date is JD 273350372.4528.} (obsId
6402), as part of the {\it Guaranteed Time Observation} (GTO)
\chandra\, program. The total effective exposure time was 88.4 ksec.
The data were acquired in {\sc very faint} mode, to ease filtering of
non-X-ray events, with six CCD turned on, the four comprising the
ACIS-I array [0,1,2,3], plus CCDs 6 and 7, part of ACIS-S. However,
data from the latter two CCDs will not be used in the following
because of the much degraded point spread function (PSF) and reduced
effective area resulting from their large distance from the optical
axis. The ACIS-I 17'$\times$17' field of view (FOV) is covered by 4
chips each with 1024$\times$1024 pixels (scale 0.49"\,px$^{-1}$). The
observation was pointed toward R.A.=10$^{\rm h}$\,44$^{\rm
m}$\,47.93$^{\rm s}$ and DEC=-59$^\circ$\,43'\,54.21", chosen to
maximize the number of stars in the FOV and close to the optical axis,
but also including most of the OB stars of the cluster. Figure
\ref{acis}-left shows \tr as seen in X-rays by our ACIS-I
observation.

Most of the observed X-ray sources in Fig.\ref{acis}-left are likely
located toward the central part of \tr, spatially constrained by the
dark V-shaped dust lane of the Carina region
\citep{1998PASA...15..202B}. However, note the small X-ray source
concentration inside this dark structure, located towards the
South-East part of Fig\,1. Deep near-IR observation are needed to
reveal counterparts of such population. We defer to a forthcoming
paper the use of some public HST\,-\,WFPCII\footnote{Hubble Space
Telescope - Wide Filed Planetary Camera 2} observations to find
signatures of star-environment interactions for some of the X-ray
sources.

\subsection{Data reduction}

Data reduction, starting with the Level 1 event list provided by the
pipeline processing at the CXO, was performed using {\sc CIAO
3.3.0.1}\footnote{http://cxc.harvard.edu/ciao/} and the {\sc CALDB
3.1.0} set of calibration files. We produced a level 2 event file
using the {\sc acis\_process\_event} CIAO task, taking advantage of
the VF-mode enhanced background filtering, and retaining only events
with grades\,=\,0,2,3,4,6 and status=0. Photon energies were corrected
for the time dependence of the energy gain using the {\sc corr\_tgain}
CIAO task. Intervals of background flaring were searched for, but none
were found. We will hereafter assume a non-variable background.  To
improve the sensitivity to faint sources, given the spectrum of the
background and that of typical sources, we filtered out events outside
the [500:8000] eV energy band.

\section{Data analysis}
\label{anal}

We built images in the three band-passes 0.5-1.5 keV (soft), 1.5-2.2
keV (medium) and 2.2-8.0 keV (hard). Before the color-coded image
combination, we corrected single band-images by variations in
exposure, sensitivity and vignetting, by computing and dividing by its
respective exposure maps. We construct a color-coded X-ray image of
the region by composition of the three soft (red), medium (green) and
hard (blue) images (see Fig.\,\ref{acis}). This image is a
17$\times$17\,arcmin field which comprises the center of Tr\,16 and
marginally the southeast part of Tr\,14. In addition to revealing a
huge number of X-ray point-like sources with different color-energies,
inspection of Fig.\,\ref{acis} suggests the presence of soft diffuse
X-ray emission in the region. In particular,
\cite{2006astro.ph..8173T} use a 57 ksec ACIS-I observation centered
on Tr\,14 to study the diffuse X-ray emission in such a region and in
the North-West part of Tr\,16. They explain the diffuse X-ray emission
as likely arising from the fast O-star winds that shock and thermalize
the surrounding medium. Because homogeneous data analysis techniques
are required to do justice to a comparison of diffuse X-ray emission
in this data-set with that of  Tr\,14, this study will be presented in
a forthcoming paper.

\subsection{X-ray source detection} \label{sect:detection}

Source detection was performed with the Palermo Wavelet Detection
code, PWDetect\footnote{See
http://www.astropa.unipa.it/progetti\_ricerca/PWDetect}
\citep{1997ApJ...483..370D}. It analyzes the data at different spatial
scales, allowing the detection of both point-like and moderately
extended sources, and efficiently resolving close source pairs. The
most important input parameter is the detection threshold (SNR), which
we establish from the relationship between background level of the
observation and expected number of spurious detections due to Poisson
noise\footnote{This last quantity was determined from extensive
simulations of source-free fields
\citep[see.][]{1997ApJ...483..350D}}.

Diffuse soft ($\sim$0.2-1.2 keV) X-ray emission has already been
identified in this region \citep{1995RMxAC...2...97C,
2006astro.ph..8173T} and causes different (non-uniform) X-ray
background levels across the FOV of our observation. The best way to
reduce the diffuse emission contribution is to discard soft photons in
the detection procedure. We compute background (BKG) levels in four
different energy ranges: 0.5-8.0 keV, 0.7-8.0 keV, 0.9-8.0 keV and
1.1-8.0 keV. Since exposure maps are needed by the source detection
algorithm, we used event files in these energy bands as input to the
CIAO tool {\sc mkexpmap} and assumed a monochromatic spectrum (kT=2.0 
keV)\footnote{http://asc.harvard.edu/ciao/download/doc/expmap\_intro.ps}.
The background level for each energy band was computed as the average
of values measured in three circular regions free of sources. If we
accept 10 spurious detections in the ACIS FOV\footnote{See reasons of
this choice in \cite{2007A&A...464..211A}.}, mean computed background
levels can be translated into different SNR thresholds for each energy
band, i.e.: 4.7, 4.65, 4.6 and 4.55, respectively. These input
parameters for PWDetect result in different numbers of detected
sources in each energy band, for instance, 1266 sources in the 0.9-8.0
keV band, greater than detected in the "canonical" 0.5-8.0 keV band
(1214 sources). This is consistent with the effect of diffuse soft
X-ray emission in masking weak sources.

A careful visual inspection was performed on the 0.5-8.0 keV and
0.9-8.0 keV source lists. We manually rejected respectively 242 and
271 detections, considered spurious either because they were produced
by different instrumental artifacts (e.g. CCD gaps, detector edges,
false detections along the read out trails), or since they resulted
from the ``fragmentation'' into discrete sources of the extended
emission making a toroidal ring around the LBV star \etacar
(Fig.~\ref{acis}). Furthermore examination of
afterglow\footnote{Afterglow is defined as the residual charge from
the interaction of a cosmic ray with the CCD. If afterglow events are
not removed from the data, they can result in the spurious "detection"
of faint sources.} contamination led us to discard 29 and 20 false
detections, respectively. The final source list was constructed by
merging both catalogues: the list of 943 sources detected in the
0.5-8.0 keV band and the 1004 sources detected in the 0.9-8.0 keV
band. Merging was performed using a criterion of maximum detection
significance. The two catalogues coincide in 798 sources, consisting
of 322 sources with Sig$_{\rm 0.5-8.0}$$>$Sig$_{\rm 0.9-8.0}$, while
for the remaining 476 sources Sig$_{\rm 0.9-8.0}$$>$Sig$_{\rm
0.5-8.0}$ (where Sig$_{x-y}$ is the detection significance in the
$x-y$ energy band). Sixty-one sources were detected only in the
0.5-8.0 keV band, while 176 are exclusive to the 0.9-8.0 keV band and
would have otherwise remained hidden because of the diffuse emission.
This procedure led to a total number of 1035 X-ray sources in the
entire field, which we analyze here.

\subsection{Photon extraction}
\label{sect:extraction}

{\small 
\begin{sidewaystable*}
\caption{\tr X-ray source catalog (see the electronic version for the
complete table).} 
\label{AEphot}
\begin{tabular}{lllllllllllccclllll}
\hline \hline \multicolumn{1}{l}{N$_{\rm x}$} &
\multicolumn{1}{l}{NAME} &
\multicolumn{1}{l}{R.A} &
\multicolumn{1}{l}{DEC.} &
\multicolumn{1}{l}{Error} &
\multicolumn{1}{l}{$\theta$} &
\multicolumn{1}{l}{Sig.} &
\multicolumn{1}{l}{Area} &
\multicolumn{1}{l}{PSF} &
\multicolumn{1}{l}{Cts} &
\multicolumn{3}{c}{Count Rates ($\times$10$^{-3}$ cnt\,s$^{-1}$)} &
\multicolumn{3}{c}{Quantiles} &
\multicolumn{1}{l}{$\overline E_{\rm x}$} &
\multicolumn{1}{l}{Var.}&
\multicolumn{1}{l}{flag.} \\
\cline{11-13}
\multicolumn{1}{l}{\#} &
\multicolumn{1}{l}{CXO\,J+} &
\multicolumn{1}{l}{[h:m:s]} &
\multicolumn{1}{l}{[d:m:s]} &
\multicolumn{1}{l}{(")} &
\multicolumn{1}{l}{($^\prime$)} &
\multicolumn{1}{l}{($\sigma$)} &
\multicolumn{1}{l}{(px.)} &
\multicolumn{1}{l}{(\%)} &
\multicolumn{1}{l}{(ph.)}&
\multicolumn{1}{l}{Tot.}&
\multicolumn{1}{c}{Soft}&
\multicolumn{1}{c}{Hard}&
\multicolumn{1}{l}{Q$_{\rm 25}$}&
\multicolumn{1}{c}{Q$_{\rm 50}$}&
\multicolumn{1}{c}{Q$_{\rm 75}$}&
\multicolumn{1}{l}{(keV)}&
\multicolumn{1}{l}{log(P$_{\rm ks}$)}&
\multicolumn{1}{l}{id}\\
\hline
   1  &  104338.73-593832.7  &  10:43:38.74  &  -59:38:32.74  &   0.57  &  10.24  &    4.62  &    1757  &  0.90  &     30  &    0.392  &    0.311  &    0.081  &  0.12  &  0.10  &  $--$ &  1.28  &  -0.38  &  -1	\\
   2  &  104341.39-594538.8  &  10:43:41.39  &  -59:45:38.81  &   0.53  &   8.56  &    6.03  &    1024  &  0.90  &     36  &    0.458  &    0.372  &    0.086  &  0.08  &  0.10  &  0.58  &  1.22  &  -0.16  &  -1	 \\
   3  &  104341.41-594224.5  &  10:43:41.41  &  -59:42:24.52  &   0.28  &   8.52  &   16.38  &     964  &  0.91  &    208  &    2.626  &    1.533  &    1.093  &  0.13  &  0.17  &  0.37  &  1.74  &  -0.39  &  -2	 \\
   4  &  104341.48-594102.2  &  10:43:41.49  &  -59:41: 2.26  &   0.43  &   8.86  &    6.15  &     997  &  0.90  &     55  &    0.704  &    0.448  &    0.256  &  0.07  &  0.10  &  0.38  &  1.25  &  -1.31  &  -2	 \\
   5  &  104343.14-594409.0  &  10:43:43.15  &  -59:44: 9.02  &   0.46  &   8.17  &    5.88  &     726  &  0.90  &     28  &    0.361  &    0.052  &    0.309  &  0.23  &  0.24  &  $--$  &  2.31  &  -0.39  &  -1	 \\
   6  &  104343.98-594655.9  &  10:43:43.99  &  -59:46:55.93  &   0.43  &   8.60  &    4.83  &     843  &  0.90  &     41  &    0.532  &    0.483  &    0.049  &  0.11  &  0.09  &  0.55  &  1.19  &  -0.76  &  -4	 \\
   7  &  104344.11-594817.9  &  10:43:44.11  &  -59:48:17.95  &   0.46  &   9.16  &    5.29  &    1042  &  0.90  &     47  &    0.602  &    0.550  &    0.052  &  0.08  &  0.07  &  0.56  &  1.01  &  -1.05  &  -1	 \\
   8  &  104345.36-593948.7  &  10:43:45.37  &  -59:39:48.78  &   0.36  &   8.89  &    7.30  &     731  &  0.89  &     54  &    0.699  &    0.498  &    0.201  &  0.11  &  0.10  &  0.31  &  1.25  &  -0.34  &  -2	 \\
   9  &  104345.37-593847.5  &  10:43:45.37  &  -59:38:47.54  &   0.40  &   9.40  &    5.83  &     970  &  0.91  &     45  &    0.580  &    0.227  &    0.353  &  0.14  &  0.11  &  0.49  &  1.36  &  -2.29  &  -1	 \\
  10  &  104345.44-594158.9  &  10:43:45.44  &  -59:41:58.95  &   0.44  &   8.11  &    5.31  &     564  &  0.90  &     25  &    0.321  &    0.150  &    0.171  &  0.14  &  0.15  &  0.43  &  1.63  &  -0.11  &  -1	 \\
  11  &  104346.39-594929.8  &  10:43:46.39  &  -59:49:29.82  &   0.37  &   9.55  &    4.72  &     474  &  0.74  &     41  &    0.634  &    0.550  &    0.084  &  0.11  &  0.10  &  0.35  &  1.22  &  -0.74  &  -1	 \\
  12  &  104348.15-594924.4  &  10:43:48.15  &  -59:49:24.41  &   0.29  &   9.32  &   12.27  &     927  &  0.91  &    132  &    1.677  &    1.276  &    0.402  &  0.12  &  0.10  &  0.35  &  1.28  &  -1.24  &  -2	 \\
  13  &  104349.40-594456.3  &  10:43:49.41  &  -59:44:56.36  &   0.28  &   7.45  &   13.68  &     418  &  0.90  &     89  &    1.138  &    1.118  &    0.020  &  0.06  &  0.06  &  0.56  &  0.98  &  -0.96  &  -3	 \\
  14  &  104350.13-594552.6  &  10:43:50.14  &  -59:45:52.69  &   0.47  &   7.54  &    5.11  &     427  &  0.90  &     15  &    0.203  &    0.248  &    0.044  &  0.08  &  0.07  &  $--$  &  1.01  &  -0.36  &  -1	 \\
  15  &  104350.71-593744.4  &  10:43:50.71  &  -59:37:44.45  &   0.42  &   9.49  &    5.21  &     905  &  0.89  &     41  &    0.539  &    0.414  &    0.125  &  0.12  &  0.11  &  0.47  &  1.33  &  -0.78  &  -4	 \\
  16  &  104350.89-595031.2  &  10:43:50.90  &  -59:50:31.25  &   0.39  &   9.76  &    5.26  &    1073  &  0.90  &     23  &    0.302  &    0.164  &    0.138  &  0.16  &  0.09  &  0.68  &  1.15  &  -2.67  &  -4	 \\
  17  &  104351.09-594024.6  &  10:43:51.10  &  -59:40:24.67  &   0.38  &   7.97  &    5.42  &     479  &  0.89  &     28  &    0.361  &    0.239  &    0.122  &  0.11  &  0.09  &  0.36  &  1.20  &  -0.31  &  -1	 \\
  18  &  104351.64-594525.1  &  10:43:51.64  &  -59:45:25.10  &   0.40  &   7.25  &    7.05  &     376  &  0.90  &     31  &    0.400  &    0.253  &    0.147  &  0.14  &  0.16  &  0.39  &  1.72  &  -3.15  &  -2	 \\
  19  &  104351.87-594035.6  &  10:43:51.88  &  -59:40:35.60  &   0.43  &   7.81  &    5.87  &     443  &  0.89  &      7  &    0.102  &    0.102  &    0.001  &  0.13  &  0.09  &  $--$  &  1.21  &  -0.34  &  -2	 \\
  20  &  104352.13-594802.0  &  10:43:52.13  &  -59:48: 2.00  &   0.40  &   8.15  &    5.23  &     528  &  0.90  &     30  &    0.388  &    0.239  &    0.149  &  0.12  &  0.09  &  0.37  &  1.16  &  -0.46  &  -4	 \\
  21  &  104352.25-594157.6  &  10:43:52.25  &  -59:41:57.68  &   0.39  &   7.28  &    5.31  &     357  &  0.89  &     25  &    0.325  &    0.301  &    0.023  &  0.07  &  0.08  &  0.93  &  1.07  &  -0.16  &  -4	 \\
  22  &  104352.48-593920.9  &  10:43:52.48  &  -59:39:20.96  &   0.23  &   8.35  &   19.14  &     592  &  0.90  &    176  &    2.250  &    1.401  &    0.849  &  0.13  &  0.16  &  0.39  &  1.70  &  -0.50  &  -2	 \\
  23  &  104354.13-594145.2  &  10:43:54.14  &  -59:41:45.24  &   0.37  &   7.11  &    4.81  &     357  &  0.90  &     25  &    0.325  &    0.101  &    0.224  &  0.09  &  0.15  &  0.38  &  1.63  &  -0.17  &  -1	 \\
  24  &  104354.20-593805.2  &  10:43:54.21  &  -59:38: 5.24  &   0.41  &   8.93  &    5.21  &     737  &  0.89  &     22  &    0.290  &    0.196  &    0.094  &  0.15  &  0.12  &  0.72  &  1.43  &  -0.52  &  -1	 \\
  25  &  104355.11-593624.2  &  10:43:55.11  &  -59:36:24.23  &   0.30  &  10.04  &   13.22  &    1205  &  0.90  &    177  &    2.246  &    1.378  &    0.869  &  0.13  &  0.15  &  0.37  &  1.59  &  -4.00  &  -2	 \\
  26  &  104355.14-594750.4  &  10:43:55.14  &  -59:47:50.42  &   0.35  &   7.72  &    7.89  &     451  &  0.90  &     44  &    0.567  &    0.470  &    0.097  &  0.07  &  0.08  &  0.50  &  1.07  &  -1.07  &  -3	 \\
  27  &  104355.47-594253.5  &  10:43:55.48  &  -59:42:53.59  &   0.29  &   6.69  &   13.04  &     278  &  0.89  &     51  &    0.664  &    0.357  &    0.306  &  0.14  &  0.17  &  0.53  &  1.74  &  -3.51  &  -2	 \\
  28  &  104355.56-594923.0  &  10:43:55.57  &  -59:49:23.07  &   0.30  &   8.57  &    8.43  &     629  &  0.90  &     77  &    0.989  &    0.807  &    0.182  &  0.09  &  0.08  &  0.35  &  1.13  &  -0.50  &  -2	 \\
  29  &  104356.25-594936.4  &  10:43:56.25  &  -59:49:36.48  &   0.33  &   8.65  &    5.36  &     648  &  0.89  &     36  &    0.471  &    0.335  &    0.136  &  0.14  &  0.09  &  0.39  &  1.21  &  -0.41  &  -1	 \\
  30  &  104356.82-594236.0  &  10:43:56.82  &  -59:42:36.03  &   0.46  &   6.57  &    4.90  &     257  &  0.89  &     12  &    0.158  &    0.139  &    0.019  &  0.12  &  0.12  &  $--$  &  1.41  &  -0.41  &  -1	 \\
\hline																					   
\end{tabular}																				   
\smallskip																				   
																					   
Notes: Column labeled with $\theta$ refers to the off-axis angle
measured in arcmin from the aim point of the observation. {\it Sig.}
is the significance of the source in number of sigma over background.
Flag Id = -1: sources detected only in the 0.9-8.0 keV band. flag Id =
-2 and -3 correspond to sources detected in both bands, but Sig$_{\rm
0.9-8.0}$\,$>$\,Sig$_{\rm 0.5-8.0}$ and Sig$_{\rm
0.9-8.0}$\,$<$\,Sig$_{\rm 0.5-8.0}$, respectively. Flag Id = -4 
refers to sources detected only in the 0.5-8.0 keV band.
\end{sidewaystable*} }

Even with the high spatial resolution of the \chandra\, ACIS-I camera,
the high source density in \tr, source photon extraction is not an
easy task. Although circular regions would contain a relatively large
fraction of the PSF for almost all source photons, the extended wings
of the PSF mean that very large regions would be needed, incurring in
the risk of contamination from nearby sources. Moreover, the resulting
inclusion of a large number of background events would reduce the
signal to noise of weak sources. On the other hand, extraction from
regions that are too small may reduce the photon statistics for
further spectral and timing analysis. To address these issues,  we
decided to use {\sc ACIS Extract} (AE) v3.79 \citep{acisextract}, an
IDL based package that makes use of
TARA\footnote{http://www.astro.psu.edu/xray/docs/TARA/}, CIAO and
FTOOLS\footnote{http://heasarc.gsfc.nasa.gov/docs/software/ftools/}
software.

This task reduces the problem of accounting for non-Gaussian shapes of
the local PSF by calculating the shape of the PSF model at each
individual source's position. For some sources the background level is
affected by the extended PSF wings of the bright sources (\etacar\,
and WR25) in the FOV. AE computes source background locally, by
defining background extraction regions as circular annuli with inner
radii 1.1 times the maximum distance between the source and the 99\%
PSF contour, and outer radii defined so that the regions contain more
than 100 ``background'' events. In order to exclude contamination of
the regions by nearby sources, background events are taken from a
``Swiss cheese'' image that excludes events within the inner annuli
radii of all the 1035 sources.

AE source extraction was performed using a PSF model that contains a
specified fraction of source events (f$_{\rm PSF}$). Generally, we
choose f$_{\rm PSF}$=90\%, and computed the contours from the PSF for
a mono-energetic source with E\,=\,1.49 keV. For 9.9\% of the sources
in the denser parts of the \tr field this fraction was reduced to
avoid contamination with other nearby sources, in the most extreme
cases down to f$_{\rm PSF}$$\sim$50\% (just 3 sources). 

Following AE science hints, we then refine the initial source
positions computed by PWDetect\footnote{PWDetect assumes a symmetric
PSF.} by correlating the source images with the model of local PSF
computed by AE libraries. This procedure was only used for those
sources lying at off-axis larger than 5 arcmin (316 sources), while
for the rest of the source (719 sources) we simply adopt mean photon
positions\footnote{Please follow Acis-Extract technical procedures at
website http://www.astro.psu.edu/xray/docs/TARA/ae\_users\_guide}. AE
also estimates local background spectra, computes redistribution
matrix files (RMFs) and auxiliary response files (ARFs), constructs
light curves, performs Kolmogorov-Smirnov variability tests, and
computes photometry in 14 different energy bands. Results of AE
procedure appears in Table\ref{AEphot}, which lists  source number in
column (1), name according to CXC naming
convention\footnote{http://cxc.harvard.edu/cdo/scipubs.html} (2), sky
position (R.A. and Dec. J2000) (3,4) with relative uncertainty (5),
off-axis angle ($\theta$) (6), significance of the detection (Sig.)
from PWDetect analysis (7), the source extraction area (8); the PSF
fraction within the extraction area, assuming E=1.49 keV (9); the
background-corrected extracted source counts in the 0.5-8.0 keV band
(NetCnts) (10); the count rates (CR = NetCnts/Exptime/PSF$_{\rm
frac}$) in three spectral bands: 0.5-8.0 keV, 0.5-2.0 keV and 2.0-8.0
keV, (11-13); source photon quantiles at 25, 50 and 75\% percent in
columns 14 to 16 (see sub-section\,\ref{sect:hardratio}), the median
photon energy ($\overline {\rm E_x}$) in (17). Column 18 is the
log(P$_{\rm ks}$) Kolmogorov-Smirnov probability of non-variability
(see section \S\,\ref{var}), and in (19) are flags from PWDetect
detection code.

\subsection{X-ray hardness ratios}
\label{sect:hardratio}

A commonly used tool to explore the spectral properties of sources
with low photon statistics is the hardness ratio 
\citep[e.g.][]{1989A&A...225...48S,2003ApJ...595..719P}. In this
conventional method, the full energy range is divided into two or
three sub-bands and the detected source photons are counted separately
in each band. Most popular definitions for a single hardness ratio
(HR) exists on basis of only two energy sub-bands: $i-$ HR=H/S or
$ii-$ HR=(H-S)/(H+S). By these definitions, HR  is very sensitive to
small changes (i.e. statistical fluctuations) in the number of photons
falling in each band. The requirement of total counts in the full
energy band is at least 40 photons (just 36\% of our sources satisfy
this constraint). Above this limit HR becomes a "reliable method" to
estimate the real hardness of sources \citep{2007A&A...464..211A}. 

\begin{figure}[ht]
\centering \includegraphics[width=8.4cm,angle=0]{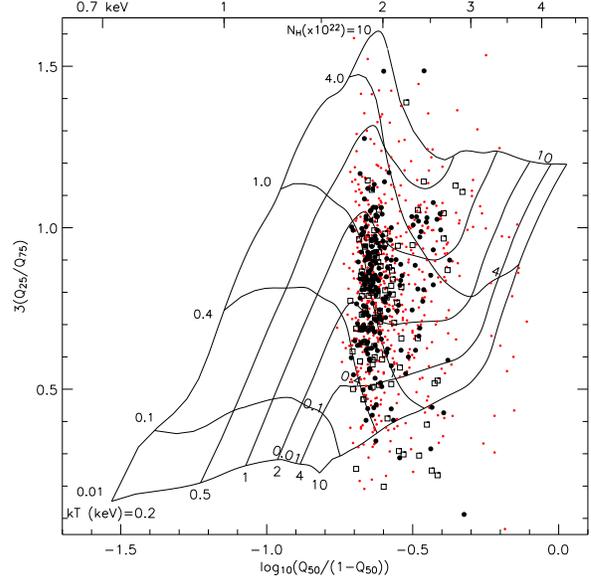}
\caption{Quantile Color-Color Diagram (QCCD). The energy scale in
the top X-axis shows the median energy values (Q$_{50}$). The grid
pattern represents the location of simulated spectra in the
diagram \citep{2004ApJ...614..508H}. Note: {\it Filled circles} and
{\it open boxes} refer to X-ray sources with and
without 2MASS counterparts (see next section), respectively.
{\it Small dots} (red) correspond to sources with unreliable quantile
values affected by poor photon statistics ($\leq$ 10 photons).}
\label{thresholds}
\end{figure}

An improved method to resolve this limitation is based on the quantile
analysis \citep{2004ApJ...614..508H}. Instead of working with
predetermined energy bands, we determine the energy E$^x$ below which
the net counts is x\% of the total counts of the source. We define
quantile Q$_x$ as: $E^x-E_{\rm min} \over E_{\rm max}-E_{\rm min}$,
where in our study E$_{\rm min}$=0.5 keV and E$_{\rm max}$=8.0 keV. We
compute median Q$_{50}$ values and quartiles Q$_{25}$ and Q$_{75}$ and
give values in Table\,\ref{AEphot}. A minor inconvenience of this
method is that for a given spectrum, various quantiles cannot be
considered independent variables, unlike the counts in different
energy bands. However, \cite{2004ApJ...614..508H} overcome this
problem by considering the log(Q$_{50}$/(1-Q$_{50}$)) vs.\
3(Q$_{25}$/Q$_{75}$) plane. Based on an extensive set of simulated
spectra, they predict loci of models in this plane. We use a set of
absorbed thermal models with plasma temperatures of 0.2, 0.5, 1, 2, 4,
10 keV and N$_{\rm H}$ equal to 10$^{20}$, 10$^{21}$,
0.4$\times$10$^{22}$, 10$^{22}$, 4$\times$10$^{22}$ and 10$^{23}$
cm$^{-2}$. Note that the spectrum changes from soft to hard as one
goes from left to right in the diagram (see Fig.\ref{thresholds}).

In a statistical sense, a comparison between sources in the grid
models suggest typical N$_{\rm H}$ and kT values distributed around 
0.8$\times$10$^{22}$ cm$^{-2}$ and 1.5 keV, but dispersed within grid 
models of N$_{\rm H}$=0.4-1.0$\times$10$^{22}$ cm$^{-2}$ and kT=1-3
keV, respectively. The absence of a systematic difference between
the position of sources with and without a 2MASS identification on the
hardness-ratio plane is likely to a combination of two observational
bias: $i$ related to the limiting magnitude of the 2MASS photometry
(K$_{\rm s}^{\rm limit}$\,$\sim$14.3 mag.); $ii$ the X-ray sensitivity
of the \chandra\, data (f$_{\rm x}^{\rm limit}$\,$\sim$10$^{-14.5}$
erg\,s$^{-1}$\,cm$^{-2}$). However is not clear, from the
observational point of view, i.e. throgouh the near-IR and X-ray flux
source distributions, that a fraction of faint stellar X-rays sources
would not have 2MASS counterparts. Deeper near-IR and X-ray
observations are needed to unveil this issue.

\section{Optical and near-IR analysis} 
\label{near-IR}

\subsection{Counterparts} \label{id} 

The large amounts of gas, dust and selective extinction of the
region, combined with the absence of available deep (V\,$>$19 mag)
optical photometry, prevent finding optical counterparts for most of
our X-ray sources. We consider it appropriate only to use optical data
to identify the massive star population (typical V\,$<$12,
\cite{1993AJ....105..980M}) of the region (i.e. Wolf-Rayet, O- and
early B-types). We use a recent compendium of massive stars in the
Carina Nebula presented by \cite{2006MNRAS.367..763S}. Out of a total
of 60 stars, 44 lie within the 0.0823 deg$^2$ FOV of our X-ray
observation. We added the binary (O5.5V\,+\,O9.5V) FO\,15
\citep{2006MNRAS.367.1450N}, not included in the list of
\cite{2006MNRAS.367..763S}. Our final list of 45 hot massive stars is
comprised of: 1 LBV, 1 Wolf-Rayet, 21 O-type stars and 22 early B-type
stars. With a cross-identification radius of 3 arcsec, a
total of 28 X-ray sources were identified. All the O-type stars were
identified as X-ray emitters, while just 7/22 ($\sim$31\%) of early
B-type (SpT. between B0V to B1.5V) have detectable X-ray emission. 
The list of massive stars with X-ray counterparts and the discussion
of their X-ray properties is presented in \S\,\ref{obstars}.

Apart from the massive stars, our X-ray population is essentially
composed of low-mass stars. In such a young cluster like Tr~16,
low-mass stars are often optically invisible, being embedded and/or
obscured by high amounts of gas and dust. We partially solve this
problem by use of the  near-IR observations, on which the impact of
dust extinction is reduced. We adopt J (1.25 $\mu$m), H (1.65 $\mu$m)
and K$_{\rm s}$ (2.17 $\mu$m) photometry from the {\it Two Micron All
Sky Survey} (2MASS) Point Source Catalog (PSC)\footnote{See
http://www.ipac.caltech.edu/2mass}. 2MASS is complete to magnitudes of
15.8, 15.1 and 14.3 mag in the J, H and K$_s$ bands, respectively. We
restricted our photometry analysis to 2MASS sources with flag quality
A, B, C or D in at least one of the three magnitudes (see explanation
in the 2MASS All-Sky Data Release User's Guide). With this restriction
20 sources were removed from our initial list of 5938 sources in the
ACIS FOV of our observation. This leaves a total of 5918 2MASS
objects\footnote{We note that the 2MASS catalog appears to have a
``hole'' around \etacar ($\sim$ 1.5 arcmin radius).}.

We began by cross-identifying our X-ray source list with the 2MASS
catalog. Identification radii, R$_{\rm id}$, were chosen to limit the
number of spurious identifications due to chance alignments, N$_{\rm
chance}$, and at the same time to include a large number of the true
physical associations, N$_{\rm true}$. Identification radii used were 
1.0, 1.5, 2.1, and 2.7 arcsec adopted for the four different off-axis
angle ranges: [0-2), [2-4), [4-7) and $>$ 7 arcmin,
respectively\footnote{The adopted radii were computed following
technical procedures presented in \cite{2007A&A...464..211A}}. Results
of the final identification are presented in \S\,\ref{near-ir} and
shown in Table\,\ref{identif}. The first seven columns are: X-ray
source number; 2MASS nomenclature for identified sources; the offset
between the two positions; J, H, K$_s$ magnitudes; photometry quality
flag (Ph.Q); confusion flag (Cont). A total of 660 X-ray sources out
of the 1035 in our list were identified with 665 2MASS objects. Five
X-ray sources (\#96, \#382, \#401, \#816 and \#1034) were
identified with two 2MASS counterparts each. After a visual
inspection, we kept only the closer counterparts. Thus the final list
of near-IR counterparts consists of 660 entries.

{\small
\begin{table*}[!ht]
\caption{Near-IR counterparts of \tr X-ray sources. 
The complete version is available in the electronic edition of A\&A.}
\label{identif}
\begin{center}
\begin{tabular}{lllccclllll}
\multicolumn{11}{c}%
{{\bfseries}}\\
\hline \hline 
\multicolumn{1}{l}{N$_{\rm x}$} &
\multicolumn{1}{l}{2MASS\,J+} &
\multicolumn{1}{l}{Offset (")} &
\multicolumn{1}{c}{J mag.}&
\multicolumn{1}{c}{H mag.}&
\multicolumn{1}{c}{K$_{\rm s}$ mag.}&
\multicolumn{1}{l}{Ph.Q} &
\multicolumn{1}{l}{Cont.} &
\multicolumn{1}{l}{A$_{\rm v}$} & 
\multicolumn{1}{l}{Mass} & 
\multicolumn{1}{l}{Notes} \\
\hline
   1 & 10433859-5938306 & 2.33 & 14.57$\pm$0.03 & 13.21$\pm$0.04 & 12.63$\pm$0.03 &  AAA &  000 &  7.93 &   2.05   & \\
   2 &       $--------$ & $--$ & $-----$ & $-----$ & $-----$ & $--$ & $--$ & $--$ & $--$ & \\
   3 & 10434145-5942245 & 0.30 & 14.18$\pm$0.03 & 13.20$\pm$0.02 & 12.82$\pm$0.02 &  AAA &  000 &  4.59 &   2.24   & \\
   4 & 10434126-5941002 & 2.62 & 15.47$\pm$$--$ & 15.31$\pm$0.14 & 14.60$\pm$0.14 &  UBB &  000 &  $--$ &   $--$   & \\
   5 & 10434295-5944080 & 1.77 & 15.63$\pm$0.08 & 14.31$\pm$0.06 & 13.27$\pm$0.04 &  AAA &  000 & 15.27 &   1.25   & K-excess\\
   6 &       $--------$ & $--$ & $-----$ & $-----$ & $-----$ & $--$ & $--$ & $--$ & $--$ & \\
   7 & 10434401-5948177 & 0.76 &  8.70$\pm$0.02 &  8.51$\pm$0.04 &  8.47$\pm$0.02 &  AAA &  000 &  1.38 &   19.2   & O9.5V\\
   8 & 10434538-5939468 & 1.92 & 15.90$\pm$$--$ & 14.88$\pm$$--$ & 14.90$\pm$0.16 &  UUC &  000 &  $--$ &   $--$   & \\
   9 & 10434536-5938471 & 0.36 & 16.02$\pm$0.10 & 14.04$\pm$0.04 & 13.02$\pm$0.04 &  AAA &  000 & 14.90 &   1.01   & \\
  10 & 10434525-5941567 & 2.61 & 14.37$\pm$0.06 & 12.99$\pm$0.05 & 12.18$\pm$0.04 &  AAA &  000 & 12.05 &   2.16   & K-excess\\
  11 & 10434659-5949292 & 1.68 & 13.69$\pm$0.05 & 12.68$\pm$0.05 & 12.25$\pm$0.04 &  AEA &  c0c &  $--$ &   2.38   & \\
  12 & 10434809-5949246 & 0.50 & 13.51$\pm$$--$ & 13.09$\pm$0.07 & 12.85$\pm$0.05 &  UAA &  0cc &  $--$ &   $--$   & \\
  13 & 10434937-5944549 & 1.45 & 12.67$\pm$0.02 & 12.20$\pm$0.03 & 12.06$\pm$0.03 &  AAA &  000 &  0.82 &   4.89   & \\
  14 & 10435007-5945530 & 0.59 & 15.56$\pm$0.05 & 14.59$\pm$0.03 & 14.19$\pm$0.07 &  AAA &  000 &  4.31 &   1.30   & \\
  15 & 10435085-5937437 & 1.28 & 14.86$\pm$0.05 & 13.74$\pm$0.05 & 13.40$\pm$0.05 &  AAA &  000 &  3.76 &   1.86   & \\
  16 & 10435088-5950307 & 0.47 & 12.20$\pm$0.02 & 11.99$\pm$0.03 & 11.88$\pm$0.02 &  AAA &  000 &  0.28 &   6.22   & \\
  17 & 10435123-5940243 & 1.06 & 14.60$\pm$0.05 & 13.38$\pm$0.04 & 12.88$\pm$0.04 &  AAA &  000 &  6.57 &   2.03   & \\
  18 & 10435132-5945239 & 2.64 & 15.54$\pm$0.08 & 14.06$\pm$$--$ & 13.55$\pm$$--$ &  AUU &  cpp &  $--$ &   1.31   & \\
  19 & 10435191-5940353 & 0.40 & 16.49$\pm$0.16 & 15.15$\pm$0.09 & 14.63$\pm$0.11 &  CAA &  000 &  $--$ &   0.70   & \\
  20 & 10435186-5948017 & 2.02 & 14.49$\pm$0.05 & 13.46$\pm$0.06 & 12.95$\pm$0.05 &  AAA &  000 &  6.60 &   2.09   & \\
  21 & 10435223-5941574 & 0.29 & 14.83$\pm$0.05 & 14.46$\pm$0.08 & 14.35$\pm$0.11 &  AAA &  000 &  $--$ &   1.88   & \\
  22 & 10435230-5939222 & 1.87 & 13.03$\pm$0.04 & 11.90$\pm$0.06 & 11.20$\pm$0.04 &  AEE &  000 &  $--$ &   3.55   & Mass-deg.\\
  23 & 10435408-5941463 & 1.19 & 15.40$\pm$0.07 & 14.24$\pm$0.05 & 13.79$\pm$0.05 &  AAA &  000 &  5.47 &   1.40   & \\
  24 & 10435419-5938073 & 2.13 & 13.03$\pm$$--$ & 13.70$\pm$0.06 & 13.53$\pm$0.07 &  UAA &  0cc &  $--$ &   $--$   & \\
  25 & 10435501-5936242 & 0.75 & 11.64$\pm$0.02 & 11.48$\pm$0.03 & 11.37$\pm$0.03 &  AAA &  000 &  0.42 &   8.28   & \\
  26 & 10435505-5947505 & 0.70 & 14.18$\pm$0.03 & 13.81$\pm$0.04 & 13.70$\pm$0.06 &  AAA &  000 &  $--$ &   2.24   & \\
  27 & 10435545-5942531 & 0.50 & 14.93$\pm$0.05 & 14.11$\pm$0.05 & 13.84$\pm$0.06 &  AAA &  000 &  2.37 &   1.81   & \\
  28 & 10435557-5949226 & 0.39 & 13.44$\pm$0.03 & 12.65$\pm$0.03 & 12.24$\pm$0.03 &  AAA &  000 &  5.08 &   2.45   & \\
  29 & 10435606-5949351 & 1.97 & 14.53$\pm$0.05 & 13.83$\pm$0.05 & 13.60$\pm$0.06 &  AAA &  000 &  1.87 &   2.07   & \\
  30 & 10435684-5942364 & 0.45 & 14.70$\pm$0.05 & 13.57$\pm$0.03 & 12.88$\pm$0.04 &  AAA &  000 &  9.63 &   1.97   & K-excess\\
\hline
\end{tabular}
\end{center}
\smallskip

Column 3 ("Offset") is the offset between X-ray and near-IR
counterpart. Ph.Q refers to the 2MASS photometric quality flags for
the J, H and, K$_s$ bands: ``A'' to ``D'' indicate decreasing quality
of the measurements, ``U'' that the value is an upper limit. The next
column refers to the contamination and confusion flag: For further
analysis we considered only sources unaffected by known artifacts,
i.e. Cont. = 000 (see 2MASS documentation for details). Masses are
given in solar units, and the last column contains information
presented in \S\,\ref{near-IR}. Note: the {\sc "Mass-deg."}
flag indicates mass degeneracy according to a Siess-based Jmag-Mass
calibration. \end{table*}}

We estimate the expected number of extragalactic sources in our
detection list by following \cite{2006A&A...455..903F} procedure. We
consider the ACIS count-rates of non-stellar sources in the {\em
Chandra Deep Field North} \citep[CDFN,][]{2003AJ....126..539A,
2003AJ....126..632B} and estimate absorption corrected count-rates
assuming N$_{\rm H}$\,=\,5$\times10^{21}$ cm$^{-2}$ (from
\Av$\sim$3.6, see \S\,\ref{near-ir}) using PIMMS and assuming
power-law spectra with index 1 and 2 \citep{2001MmSAI..72..831G}.  We
then compare these count rates with upper limits taken at random
positions in the ACIS FOV. For $\Gamma$ between 1 and 2 we obtain 72
to 95 expected extragalactic sources. Given the intrinsic near-IR
fluxes of these sources and the absorption toward \tr, they are
expected to be among the 385 without NIR counterparts (cf. Flaccomio
et al. 2006). This means that no more than 18 to 24\% of the
unidentified X-ray population is related to extragalactic sources.

\subsection{Unidentified X-ray sources}
\label{un-id}

A large population of young stars, proto-stars, deeply embedded in
dense circumstellar gas and dust should be present in the Carina
Nebula \citep{2003ApJ...587L.105S}. However optical and near-IR
counterparts of YSOs are difficult to detect. Fortunately, X-ray
emission is expected in young stellar objects along all their initial
phases \citep{2000ApJ...532.1097M}. X-ray photons easily escape from
dense circumstellar material, where absorption process becomes
important mainly for energies below 1.2 keV
\citep{1983ApJ...270.119M}. This makes hard X-ray energies the most
appropriate "window" to detect counterparts of deeply embedded young
sources.

Of the 375 X-rays sources without near-IR counterparts, just a small
fraction ($\sim$20\%) is expected to be extra-galactic contamination
(see \S\ref{id}). We are thus dealing with about $\sim$300 candidate
young (first stage) low-mass stars, highly obscured by circumstellar
material. In our data these sources typically have lower X-ray photon
statistics than those with near-IR counterpart, i.e. $\sim$8 vs.\ 28
average photons, respectively. In X--rays, no quantitative differences
in the median energy and spectral quantiles were found for the X--ray
sources with and without near-IR counterparts.

\subsection{Near-IR properties of identified X-ray sources}
\label{near-ir}

We now investigate the near-IR properties of the X-ray sources. For
this purpose we restrict our analysis to sources with high quality
photometry ({\em Ph.Q}=$AAA$) and no confusion ({\em Cont.}=000). With
these requirements the total number of IR sources in the ACIS FOV is
reduced from 5918 to 2178. We have also set a further requirement on
near-IR counterparts of X-ray sources, that their J, H, and K$_{\rm
s}$ magnitude errors be all $<$0.1 mag. All these requirements yield
367 X-ray sources with good near-IR counterparts, out of the original
660.

\begin{figure}\includegraphics[width=8.8cm,angle=0]{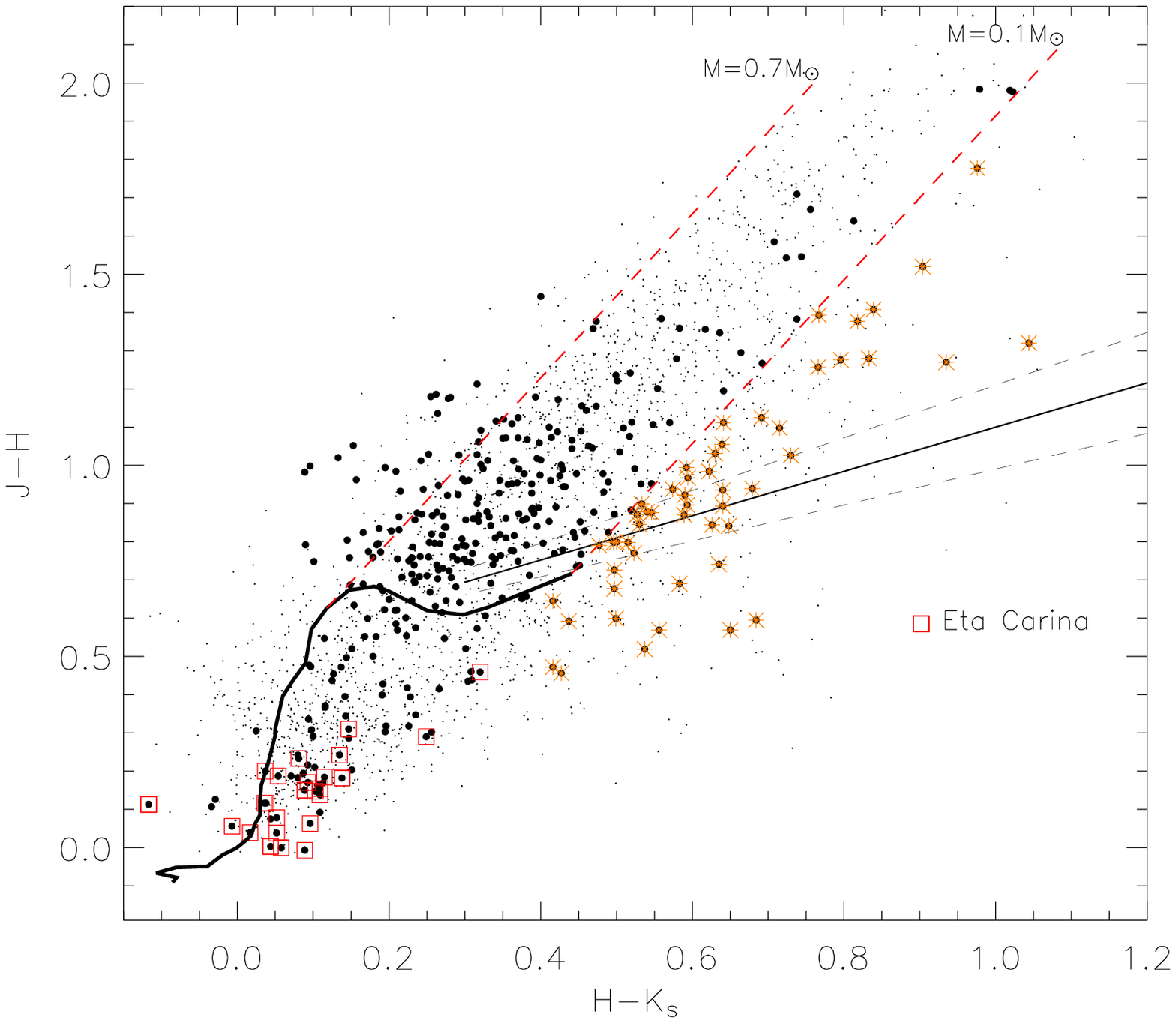}
\caption{JHK$_{s}$ color-color diagram from high quality 2MASS
photometry. {\sc Filled circles} and {\sc dots} refer to X-ray
detected and undetected 2MASS sources, respectively. {\sc Asterisks}
show the source population with intrinsic K$_{\rm s}$ excess.
Identified 2MASS source known as O- and early B-type stars are
indicated by {\sc squares}. The main sequence is shown for reference.
We also show the CTTS locus of \citet{1997AJ....114..288M} and
reddening vectors ({\it dashed lines}) with length corresponding to
\Av\,=\,13 mag. Note:  $i-$: the 2MASS photometry of \etacar is
severely affected, with {\em Phot.Qual.}='DDD'; $ii-$: the peculiar
position of the {\em red box} with H-K$_{\rm s}<$-0.1 corresponds to
the massive binary CPD-592628AB (SpT: O9.5V+B0.3V).} \label{jh_hk}
\end{figure}

\begin{figure*}[!ht]
\includegraphics[width=9.05cm,angle=0]{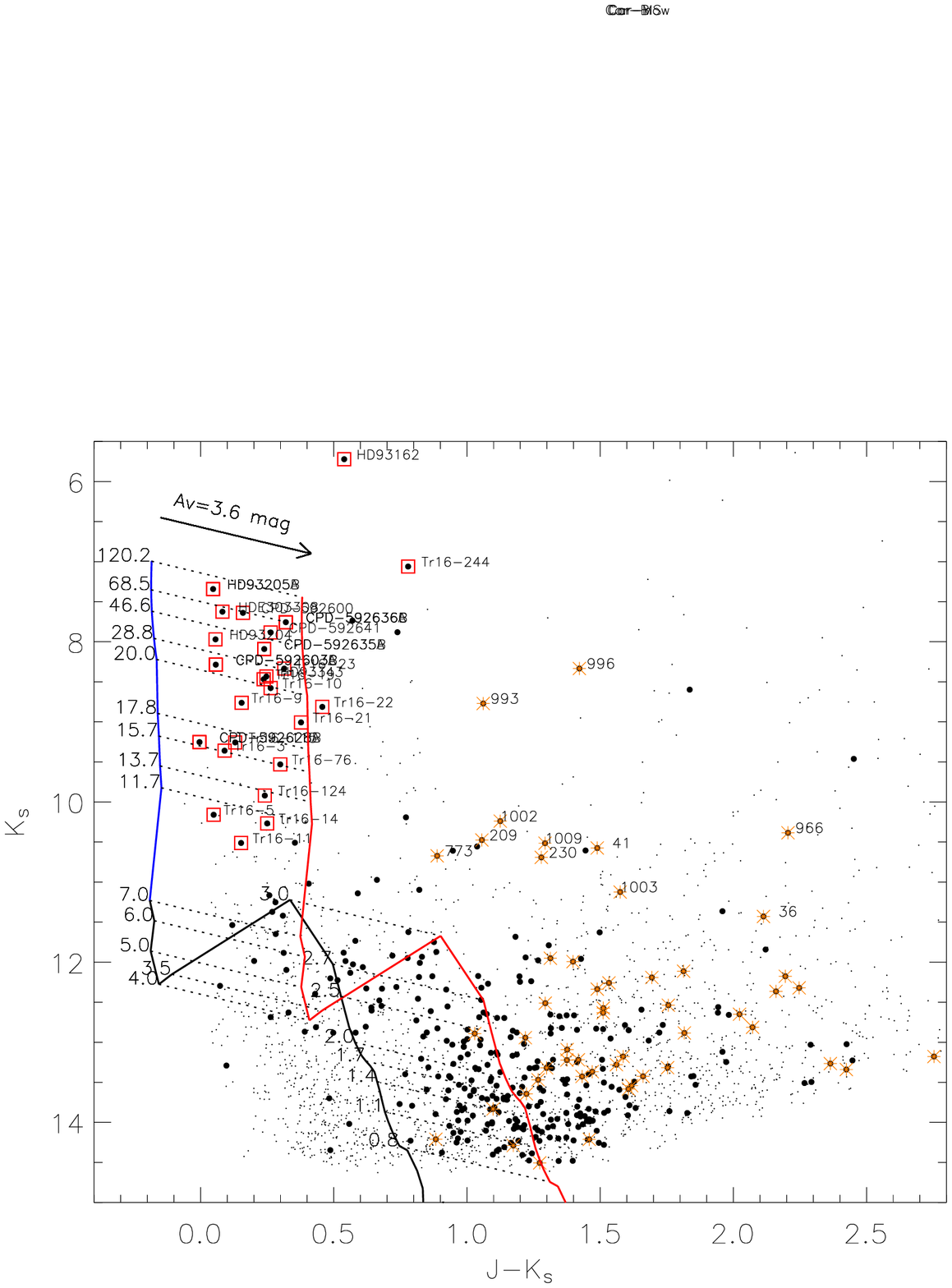}
\includegraphics[width=9.05cm,angle=0]{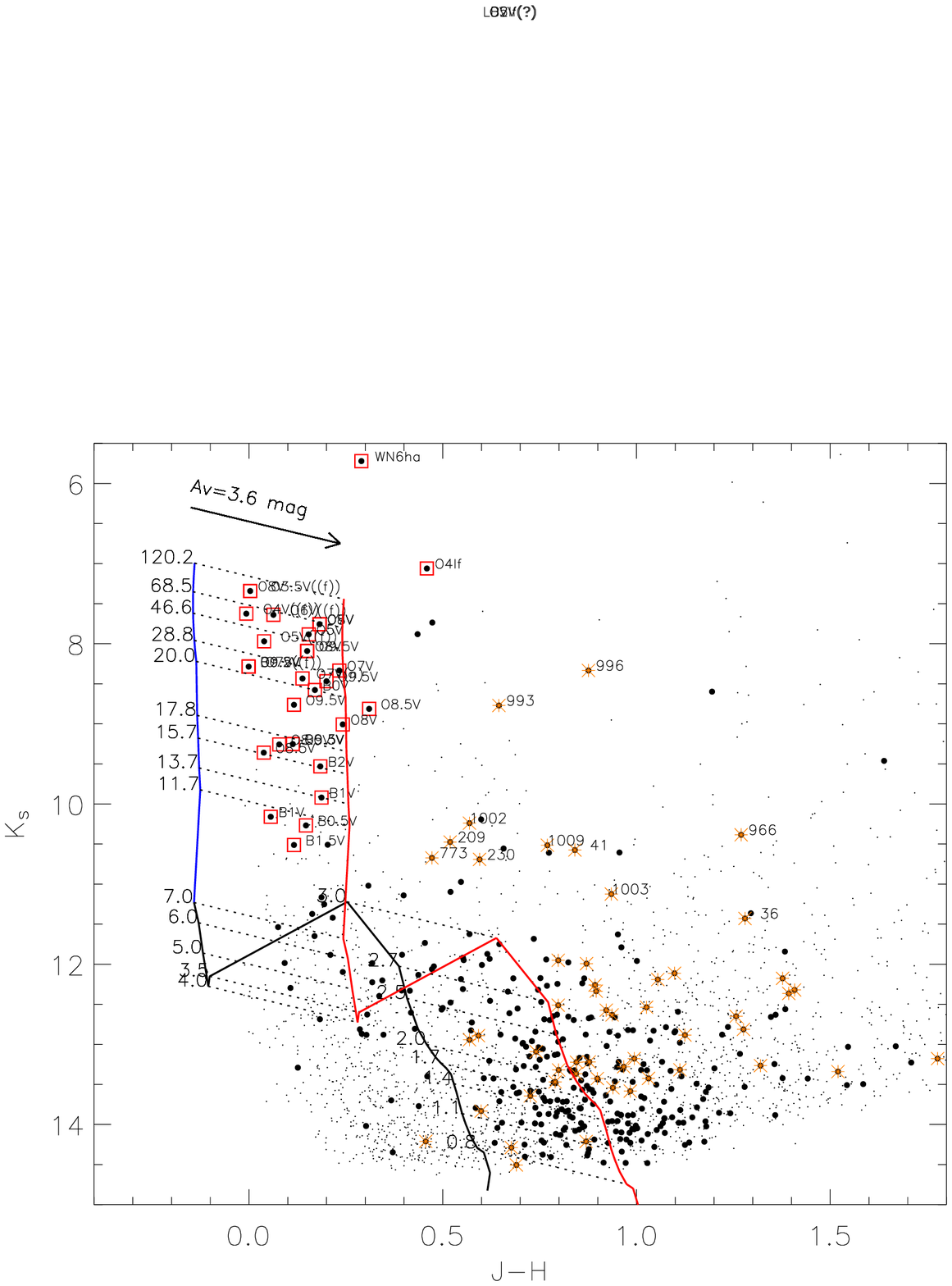}
\caption{CM diagrams of the \tr region. Symbols as in figure
\ref{cmdiag}. The two parallel curves indicate the expected cluster
loci for the assumed distance and no reddening, and for the mean
reddening \Av=3.6 mag. Masses are indicated on the left side of the
unreddened isochrone. Identified OB stars show K$_{\rm s}$ magnitudes
below 11. Stars with K$_{\rm s}$ excess show typically J-K$_{\rm
s}>1$. Note: The limitation imposed by the 2MASS photometry (K$_{\rm
s}\sim$14.5 mag) corresponds to a mass limit of 1.1 M$_\odot$. Labels
near OB stars refer to {\it Simbad names} ({\it left}) and respective
{\it Spectral types} ({\it right}). The fact that stars Tr16\,-\,244
(O4If) and HD93129AB (WN6ha+O4f) appear above the 120 M$_\odot$ track
does not mean that they are more massive, since their evolutionary
stages cannot be compared with a MS isochrone. \etacar was discarded
because of the bad 2MASS photometry (Phot.qual.='DDD', see text).
Sources labeled with numbers indicate stars (with masses above 12
M$_\odot$) probably surrounded by disks.}\label{cmdiag} 
\end{figure*}

Figure\,\ref{jh_hk} shows the J-H vs. H-K$_{\rm s}$ color-color (CC)
diagram for these AAA-flagged sources. We also plot for comparison the
MS \citep{1995ApJS..101..117K}, the Classical T-Tauri Stars (CTTS)
locus of \citet{1997AJ....114..288M}, and reddening vectors starting
from these loci and with slope ($\rm A_{Ks}/E(H-K_s)$=0.125)
corresponding to the extinction law given by
\cite{2003ApJ...597..957H}. \tr members with purely photospheric
emission should lie in this reddening band. Otherwise, Young stellar
objects (YSOs), such as Classical T\,Tauri and Herbig Ae/Be stars,
because of the NIR excess emission originating in the inner parts of
their circumstellar disks, are often found to the right of this band,
i.e. in the CTTS locus. Fifty-one (out of 367) X-ray sources, i.e.
likely \tr members, have colors consistent with the (reddened) CTTS
locus. This  means a fraction of 51/339 (28 OB stars were discarded)
$\sim$\,15\% of all (low-mass) identified X-ray sources in the CC
diagram. Of all 51 sources with intrinsic K-excess i.e. disk-star
systems, eleven (Src-Id: 36, 41, 209, 230, 773, 966, 993, 996, 1002,
1003 and 1009) appears below the CTTS vectors, but with intrinsic
bright K$_{\rm s}$ magnitudes, as is shown in both panles of
Fig\,\ref{cmdiag}. They are probably intermediate- to high-mass young
stars with an intrinsic K-band excess that would be produced by
massive accretion disks and/or extended envelopes surrounding massive
YSOs. If confirmed, they will contribute for about 40\% of the total
massive star population of the \tr region.

Figure\,\ref{cmdiag}-left shows the K$_{\rm s}$ vs. J-K$_{\rm s}$
color magnitude (CM) diagram for the same stars plotted in Fig.
\ref{jh_hk}. We also show for reference the expected cluster locus:
the intrinsic K$_{\rm s}$ magnitudes and J-K$_{\rm s}$ colors for
stars earlier than B5V were taken from the MS calibration of 
\citet{2000A&A...360..539K} and \citet{1989LNP...341...61B}. For later
spectral types (masses between 0.1 and 7 M$_\odot$), we adopted the
3\,Myr isochrone from \cite{2000A&A...358..593S}, converted to the
observational plane using the calibration given by
\citet{1995ApJS..101..117K}. The adopted MS and 3\,Myr isochrone
overlap satisfactorily.

In order to estimate the typical visual absorption of cluster members,
we computed the distance of each X-ray source from the cluster locus
along the reddening direction on the K$_{\rm s}$\,vs\,J-H plane (see
Fig.\,\ref{cmdiag}-right). Resulting \Av values for individual
sources\footnote{Individual photometric errors of up to 0.1 mag at
K$_{\rm s}$ and J-H color errors of $\sim$0.14, could results for \Av
errors up to $\sim$0.7 mag.} are listed in column 9 of
Table\,\ref{identif}. Note that for 11.2$<$K$_{\rm s}<$12.8 the
absorption cannot be constrained because the reddening vector
intersects the cluster locus more than once. In both CM diagrams, some
X-ray sources, of the order of 20, lie to the left of or relatively
close to the unreddened cluster loci. These are likely to be
foreground MS stars and thus their \Av values either cannot be
computed or are close to zero. In a statistical sense,  the median \Av
value of OB stars (\Av=2.0$\pm$0.8 mag.) is lower than that computed
for low-mass stars (\Av=3.6$\pm$2.4 mag., considered to be the typical
absorption of the cluster). Obviously, the above estimates depend on
the reliability of the assumed cluster locus and on the assumption
that disk-induced excesses do not significantly affect the J and H
magnitudes. To caution of possible contamination and/or anomalous J
magnitude, but in particular for the H-band, the median \Av for the
low-mass star population was estimated by discarding sources with
intrinsic K$_{\rm s}$ excesses (i.e. labeled with asterisks in figures
\ref{jh_hk} and \ref{cmdiag}). We must note that: $i-$ the dispersion
along the J-K$_{\rm s}$ axis (see Fig\,\ref{cmdiag}-left) indicates
differential absorption of the region, and translates in a broad \Av
distribution, with a spread 1$\sigma \sim$2.4 mag. $ii-$ the
difference between median \Av values of low-mass and OB stars suggests
a clearing effect of strong winds and radiation field of massive stars
on their surrounding environment. This conclusion was also reached by
\cite{2007A&A...464..211A} for massive stars of the Cyg\,OB2 region.
However, compared to the Cyg\,OB2 region, the fraction of disk-stars
members in the \tr region is about four times larger. 

Finally, we use the 2MASS J-band magnitudes to obtain an estimate of
stellar masses for 510 counterparts (of a total 660 identified stars)
with J-band {\em Phot.Qual.}='A' to 'D'. We compute the mass vs.
J\,mag relationship appropriate for the cluster mean age (3\,Myr),
distance (DM=11.78 mag) and extinction (\Av=3.6 mag), this latter
obtained as described above for the cluster locus in the CM
diagrams\footnote{The choice of the J band is justified because (i) in
the presence of disk excesses the J-band is the most representative of
the photospheric emission i.e. least affected and (ii) the mass ranges
in which the mass-luminosity relationship is degenerate are narrower
than for a similar relationship in the H and K$_{\rm s}$ bands.}. We
use \cite{2000A&A...358..593S} models to compute PMS tracks of low-
and intermediate-mass stars (masses $\leq$ 7 M$_\odot$). We
interpolated the J\,mag - mass relation, using the J magnitude vs.
mass relationship at 3\,Myr (see column 10 of Table\,\ref{identif}),
to compute individual masses of stars\footnote{This method suffers of
photometry inaccuracy, distance and age spread. Mass values should be
adopted carefully and should not be considered to measure the slope of
the mass function.}. Unfortunately, the relation mass-J\,mag
degenerates in the mass ranges 0.2-0.65\,M$_\odot$ (3 sources) and
2.7-4.52\,M$_\odot$ (21 sources). We indicate these sources with
"Mass-deg." flag in column 11 of Table\,\ref{identif}, and give mean
mass values for sources lying in these two ranges, i.e. 0.42 and 3.55,
respectively. Computed masses over 7 M$_\odot$ are potentially
affected by large uncertainties in the extrapolated J\,mag - mass
relation, and these values were excluded from further analysis. In
Table \ref{identif} we give masses for a total of 510 stars, of which
410 range between 0.65 and 2.52 M$_\odot$.

\section{X-ray variability} \label{var}
\begin{figure*}[!ht]
\includegraphics[width=6cm,angle=0]{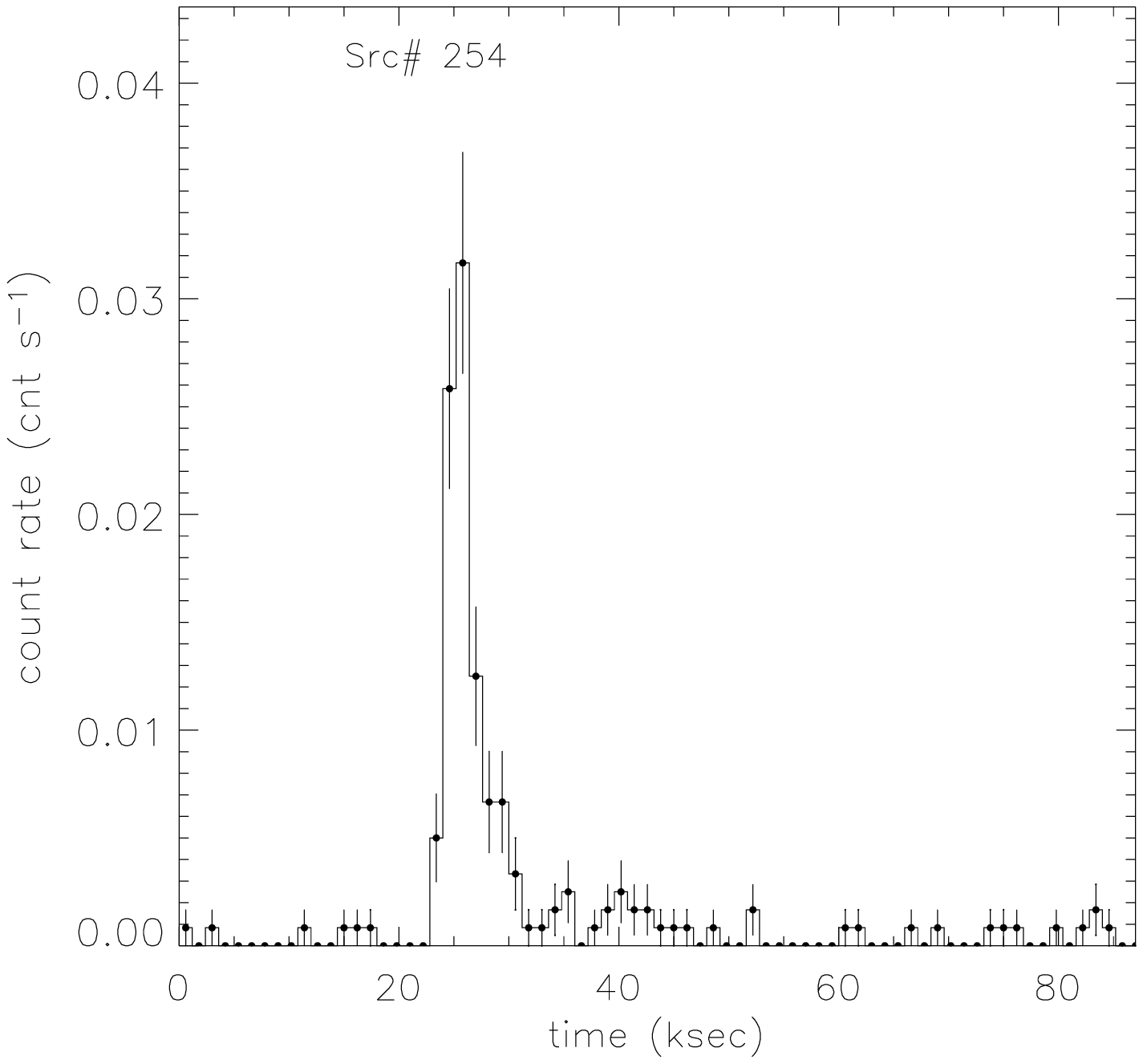}
\includegraphics[width=6cm,angle=0]{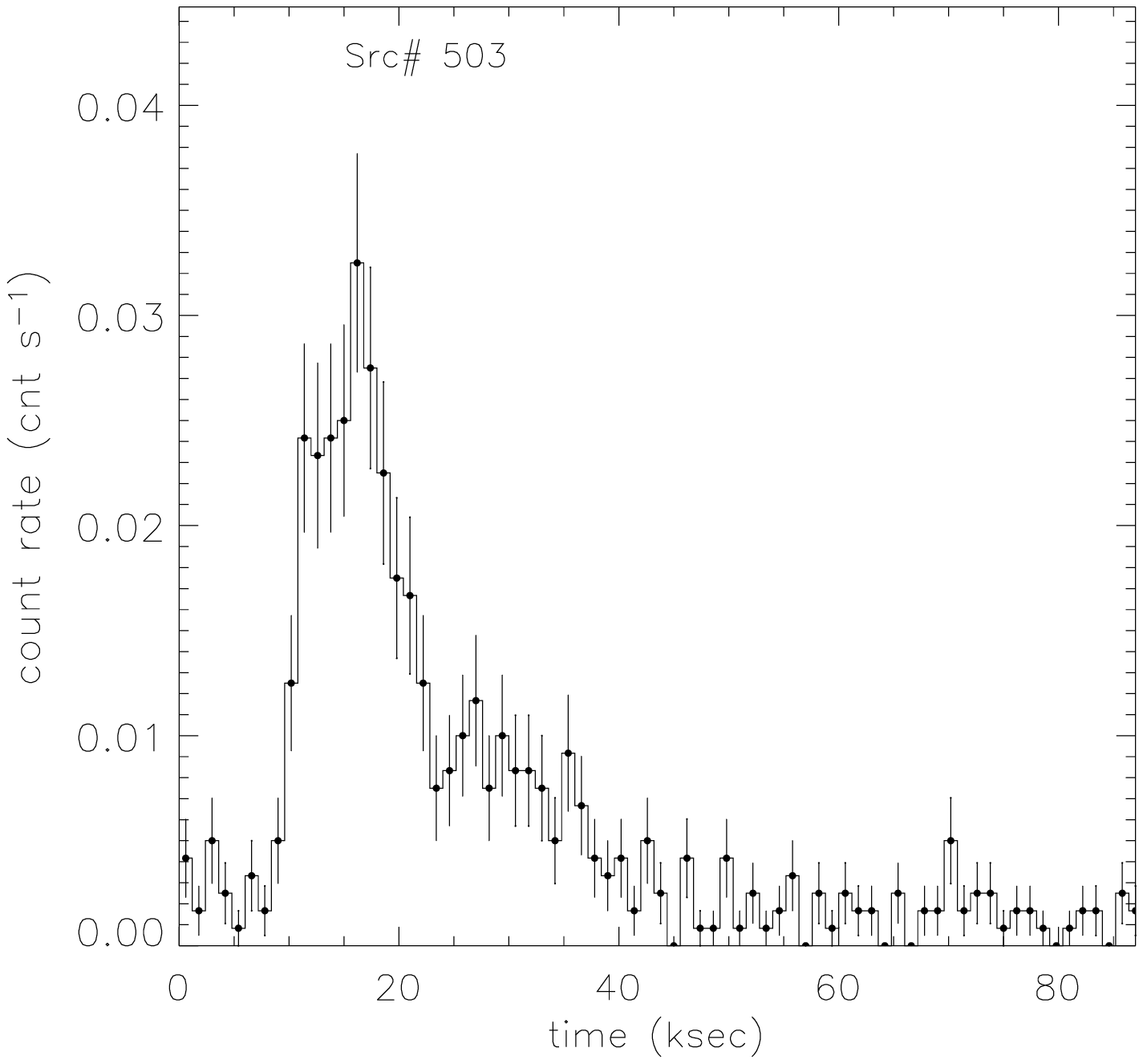}
\includegraphics[width=6cm,angle=0]{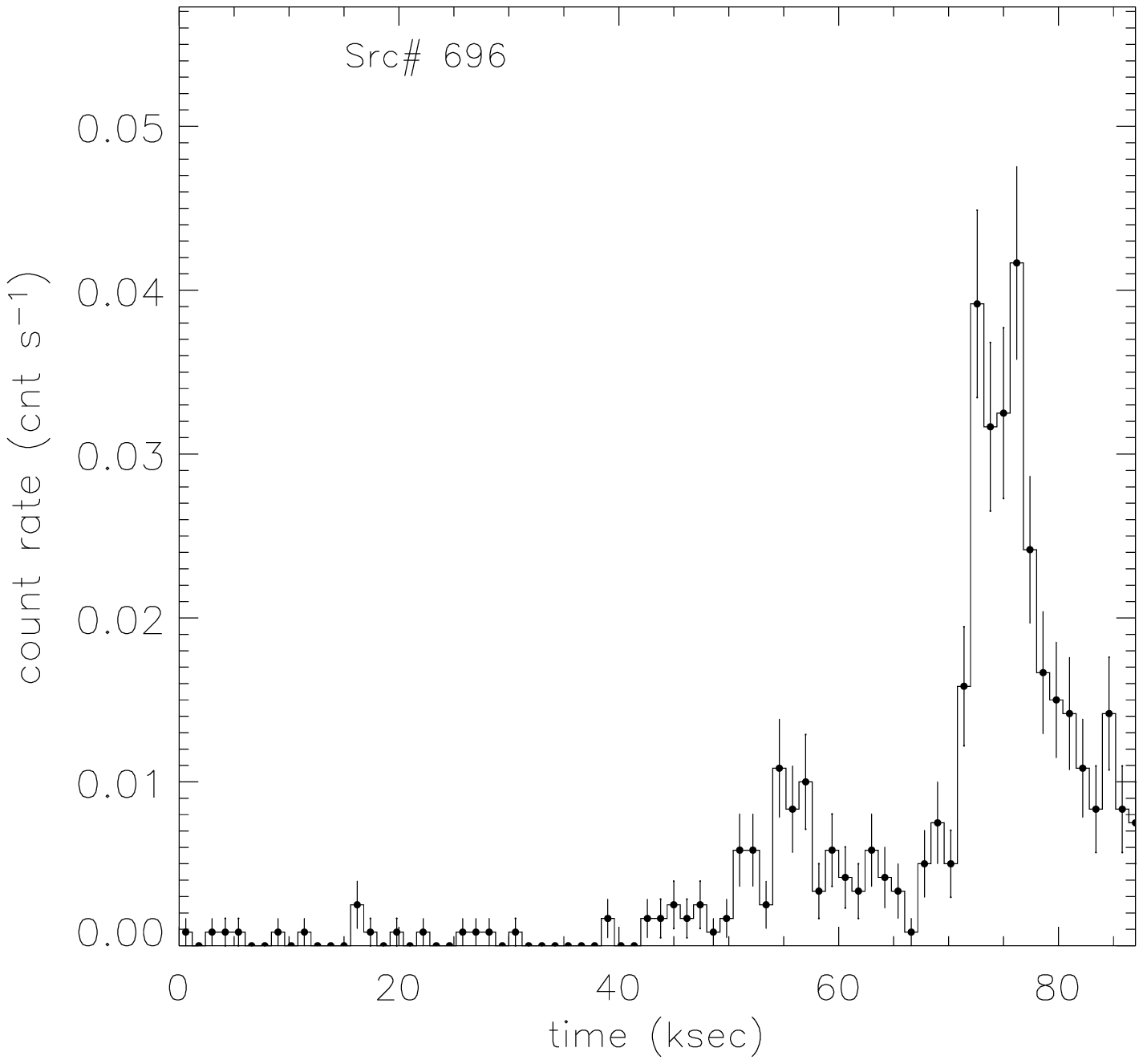}\\
\includegraphics[width=6cm,angle=0]{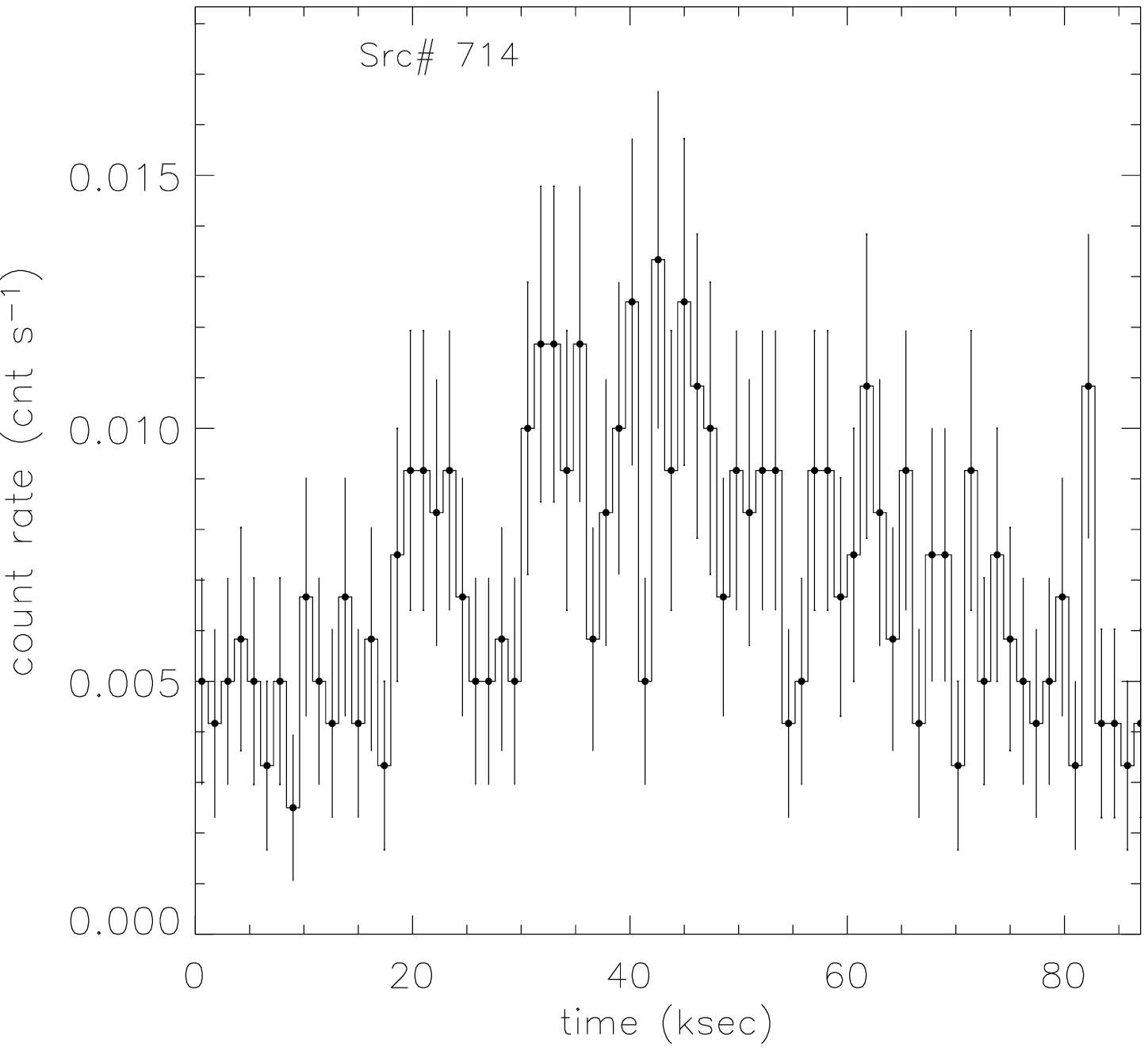}
\includegraphics[width=6cm,angle=0]{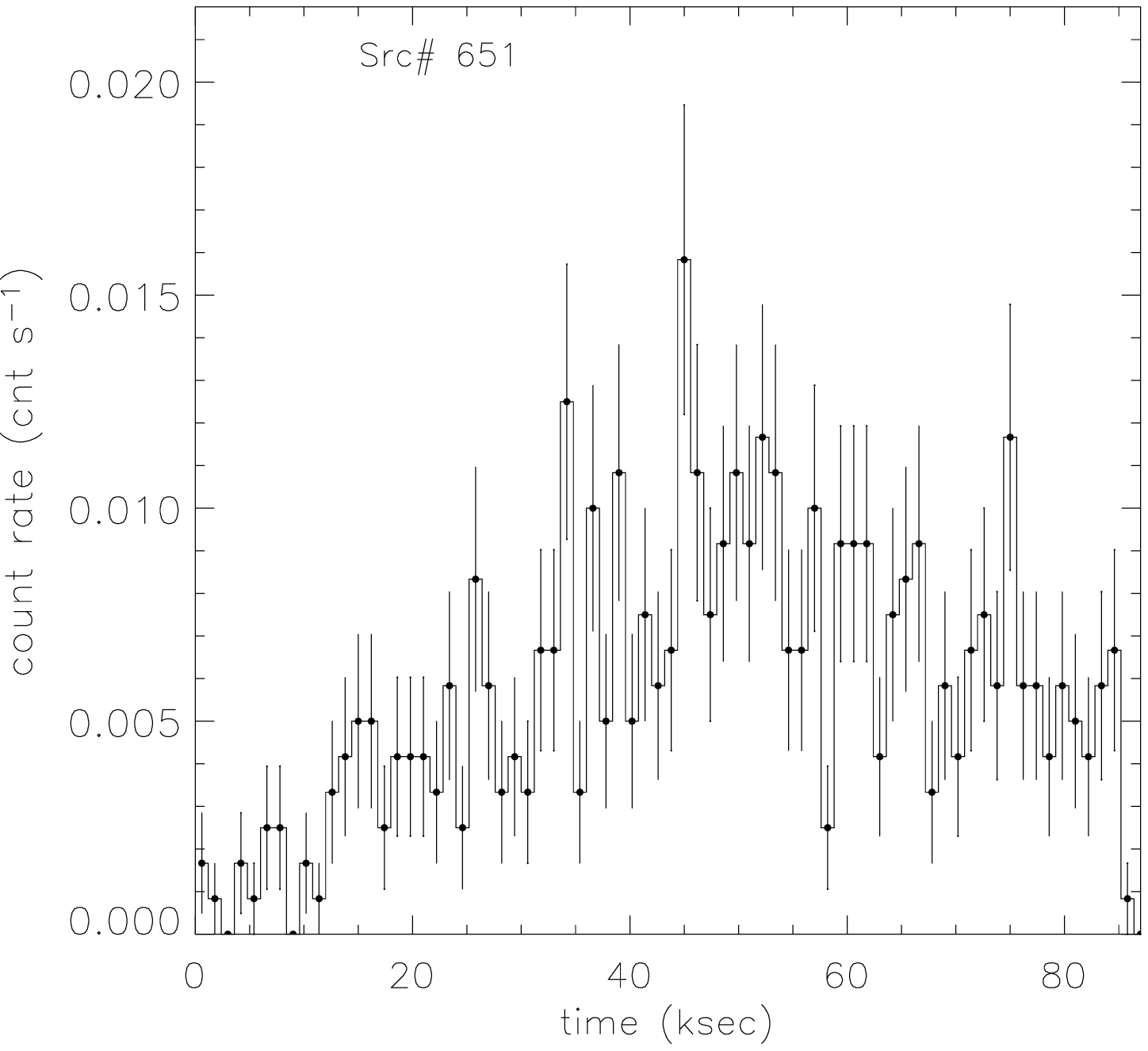}
\includegraphics[width=6cm,angle=0]{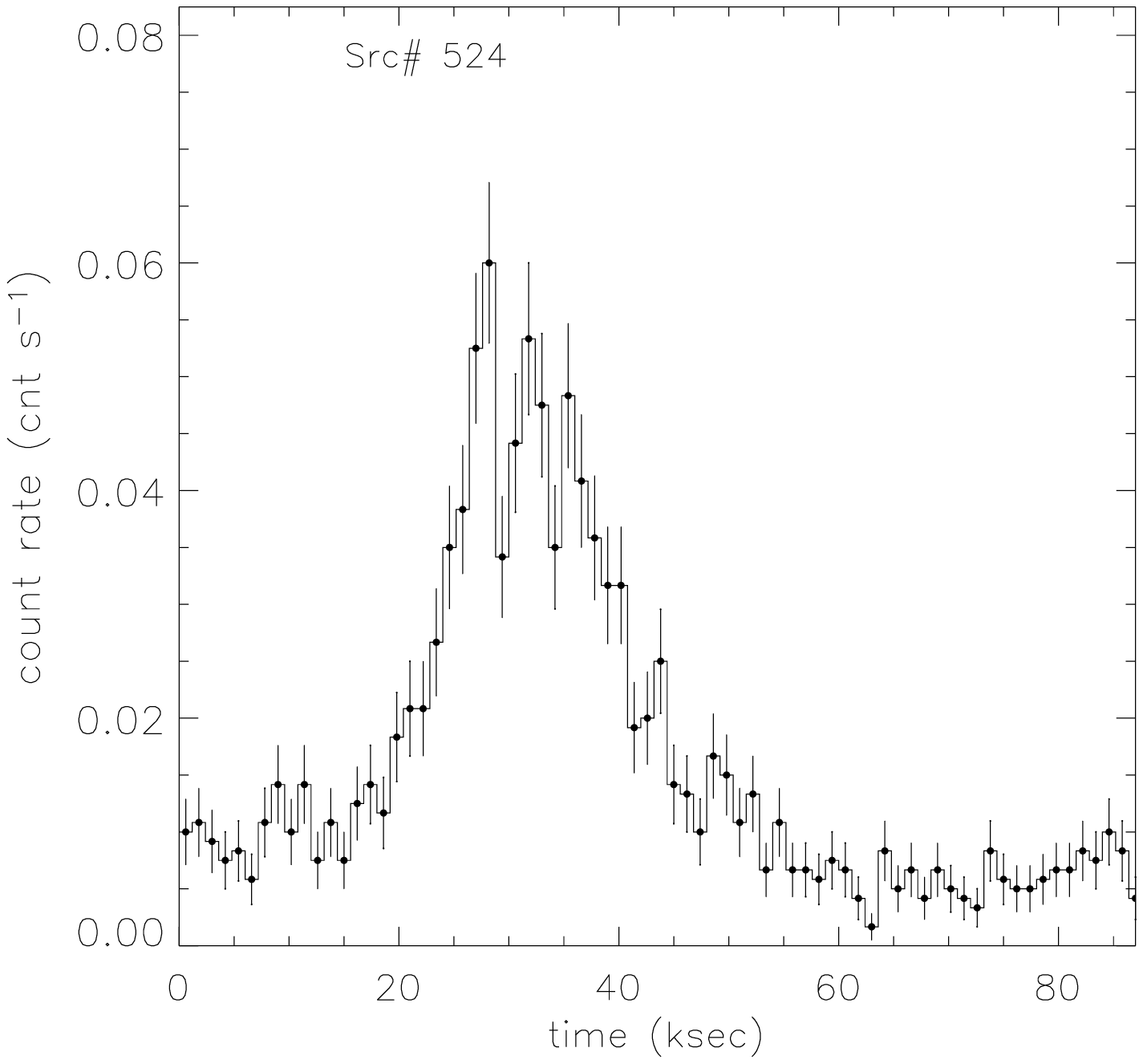}
\caption{Light curves (in the 0.5-8.0 keV band) showing different
variability scenarios occurring among our 77 variable sources. Bin
size is 900 sec. The source number is indicated in each panel. Upper
panels show a flare-like behavior with very quick rises and decay
phases of hours, while bottom panels are slowly modulated variability.
A probable flare+modulation variability is occurring in the last
panel.}
\label{ltc}
\end{figure*}

PMS stars have high levels of X-ray activity that are commonly
attributed to a "scaled up" solar-like corona formed by active
regions. X-ray variability over a wide range of time scales is common
in all magnetically active stars
\citep[e.g.][]{1999ARA&A..37..363F,2003SSRv..108..577F,2004A&ARv..12...71G}.
On long time scales, this includes rotational modulation of active
regions, their emergence and evolution and magnetic cycles
\citep[e.g.][]{2003A&A...407L..63M,2005ApJS..160..450F}. Most of the
observed variations have short time-scales ($\sim$ hours), however, and
can be attributed to small scale flares triggered by magnetic
reconnection events.

We first investigated X-ray variability in our sources using the
non-binned one-sample Kolmogorov-Smirnov (KS) test
\citep{1992nrfa.book.....P}. This test compares the distribution of
photon arrival times with that expected for a constant source. The
test was applied to photons in the source extraction regions, which
also contain background photons. Given that the background was found
to be constant with time (\S 2.1), the results, i.e. the confidence
with which we can reject the hypothesis that the flux was constant
during our observation, can be attributed to the source photons.
Table\,\ref{AEphot}, column 18, reports the logarithm of the KS-test
significance with values $<$\,-4 truncated at that value: sources with
log(P$_{\rm KS}$)\,$<$\,-3.0 can be considered almost definitely
variable as we expect at most one of the 1035 sources (i.e.
$\leq$0.1\%) to be erroneously classified as variable. Seventy-seven
X-ray sources  ($\approx$7.4\% of the total) fall in this category.
Fifty-five sources with -2.0\,$<$\,log(P$_{\rm KS}$)\,$<$\,-3.0 can be
considered as likely variable, although about half of them suffers of
low photon statistics. These numbers of sources are lower limits to
the total number of variable sources in the region for several
reasons: i) most of the observed variability is in the form of flares,
i.e. events that are shorter than our observation and with a
duty-cycle that  may be considerably longer
\citep{2005ApJS..160..423W}; ii) the sensitivity of statistical tests
to time variability of a given relative amplitude depends critically
on photon statistics (see \cite{2007A&A...464..211A}). Hereafter, we
consider variable those 77 sources with log(P$_{\rm ks}$)$<$-3\,. 

To get a more accurate description of the detected variability, we
extracted binned light-curves for each of the 77 variable sources in
the region. We adopt a bin length of 900 seconds, a compromise between
bins that are long enough to reach a good signal-to-noise ratio per
bin for most sources and sufficiently short to resolve the decay phase
of typical flares. Since the background of our observation is both low
(negligible for many sources) and constant in time, we did not apply
any background subtraction to the presented light-curves. In Fig\,
\ref{ltc} we show examples of the different behavior among the
light-curves of variable sources. Source\,\#254 like others (\#35,
\#270, \#438, \#468, \#480, \#596, \#761 and \#975) experience
``impulsive'' flares with very quick rises and decay phases of only a
few hours. Others (sources\,\#503 (\#36, \#41, \#136, \#164,
\#260, \#489, \#523, \#731 and \#811) show longer (2 to 10 hours) flares.
In several instances a second impulsive event is visible during the
exponential decay of a previous flare (e.g. sources \#36, \#52, \#87,
\#196, \#564, \#600, \#620 and \#623). The case of source\,\#696 is a
combination of both variability types, with two consecutive flares.
Other sources like \#714 and \#651 (e.g. \#27, \#251, \#271, \#793,
\#839, \#890, \#904 and \#980) show light-curves that, instead of
showing typical flares, are characterized by slow continuous rises or
decays that might be explained by rotational modulation of
non-homogeneously distributed plasma \citep{2005ApJS..160..450F}.
Finally, light-curves, like those observed for sources\,\#524, \#71
and \#933, seem to be related to a combination of flare like activity
and rotational modulation.  

\subsection{Variability in massive stars}

Because X-ray emission from O stars, which is believed to be unrelated
to solar-like magnetic activity, comes from the integrated emission
from a large number of small shocks randomly occurring in their strong
winds \citep{1997A&A...322..878F, 1999ApJ...520..833O}, on average,
global X-ray variability is not expected to occur. However, it is
surprising that three (out of 28) massive stars, namely Tr\,16-11
(B1.5\,V, source\,\#136), Tr\,16-5 (B1\,V, source\,\#489) and the binary
HD\,93205 (O3.5V((f))+O8V, source\,\#242), are significantly variable,
with log(P$_{\rm KS}$) values lower than -3. The origin of the
observed flare-like variability in the first two sources (B-type
stars) is probably coronal activity of unresolved late-type
companions. This hypothesis was also proposed for Tr16-11 itself by
\cite{2003ApJ...589..509E}. 

\begin{figure}[!ht]
\includegraphics[width=8.3cm,angle=0]{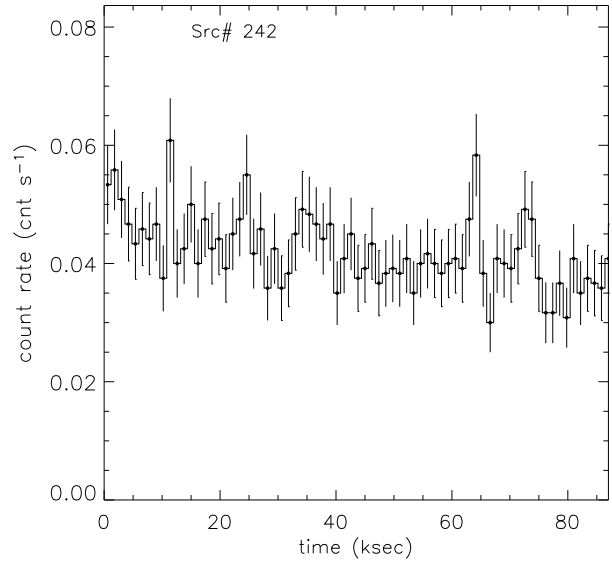}
\caption{X-ray light-curve of the massive binary (O3.5V((f))+O8V)
HD\,93205. This light-curve shows: $i-$ a uniform decay probably
related to the orbital motion of the system, with changing absorption
N$_{\rm H}$ of the colliding-wind region along the line of sight, plus
$ii-$ a probable short-term variability (see text).} \label{93205}
\end{figure}

Fig\,\ref{93205} shows the X-ray light-curve of the HD\,93205 binary
system. Two different processes may be acting simultaneously to
explain the observed variability:  \begin{itemize} \item [$i-$] The
decrement of count rate agrees with phase-locked X-ray variability
reported by \cite{2001MNRAS.326...85M}. The X-ray count rate decreases
from $\sim$ 0.055 to 0.033 cts/s (about 35\%) in about 88 ksec
($\sim$1 day) of continuous observation, i.e. about 16\% of the
orbital period. This is in agreement with X-ray emission from
colliding winds contributing most of the total detected emission. 
However, we cannot discard that magnetically channeled wind streams,
at the O3.5\,V((f)) primary, collide with the cool and dense postshock
plasma at the magnetic equator
\citep[e.g.\,$\theta$\,Orionis-C,][]{2005ApJ...628..986G}. A detailed
spectral and time (phase-resolved) study should be follow to discern
between these two possibilities. \item [$ii-$] The observed short time
X-ray variability (log(P$_{\rm KS}$)=-3.07). While dynamical
instabilities \citep{1990ApJ...362..267L}\footnote{Whenever the wind
velocities of two stars are not equal, shocked CWR should be subject
to the rapid growth of dynamical (Kelvin-Helmholtz) instabilities,
reaching the radiative cooling state. This limit becomes important for
massive binaries with typical orbital periods $\leq$20 days
\citep{2004ApJ...611..434A}} in the Colliding Wind Region (CWR) seems
to be more appropiate, the magnetic reconnection from an unknown
low-mass tertiary companion\footnote{Variability like HD 93205's has
been seen also on theta 2\,Ori-A and has been interpreted by
\cite{2006ApJ...653..636S} as evidence of binary-induced reconnection.
However, Image reconstruction shows that hard, short-term variations
are from an intermediate-mass tertiary companion at $\sim$0.3 arcsec
\citep{2002ApJ...565.1216H}.} cannot be rule out.\end{itemize}

We remark that the origin of X-ray variability in high mass stars is
out of the scope of this work, it needs for a large X-ray data-set. We
leave this subject to a further investigation.

\section{Spectral analysis}
\label{xspec}

In order to characterize the hot plasma responsible for the X-ray
emission of \tr\, stars, and to estimate their intrinsic X-ray
luminosities, we analyzed the ACIS spectra of the 615 (out of all
1035) sources with more than 20 net photons (NET\_CNTS), i.e.
corrected for local background. Spectral parameters for sources with
less than 20 net photons are too much ill-constrained
\citep{2007A&A...464..211A}, and thus were not determined. Moreover,
high local background could affect the reliability of computed
spectral parameters. We define the fraction f$_{\rm cont}$ as the
ratio between BKG\_CNTS (number of background counts in the source
extraction region) and NET\_CNTS. We accepted spectral fits for
sources with f$_{\rm cont}$$<$1 and NET\_CNTS$\geq$20 photons. Of all
1035 sources, only 563 satisfy both conditions above, while 119 show
f$_{\rm cont}$$<$1 and 353 lie in the low statistics regime (i.e.
NET\_CNTS$<$20 ph).

Source and background spectra in the 0.5-8.0 keV band were produced
with AE (see \S \ref{sect:extraction}), along with individual
``redistribution matrices files'' (RMF) and ``ancillary response
files'' (ARF). For model fitting, spectra were grouped so to have a
specified number of events in each energy bin. Grouping was tuned to
the source statistics and we chose 2, 5, 7, 10, and 60 counts per
channel for sources with net-counts in the following ranges: [20-40],
[40-100], [100-200], [200-500], and [500-**]. Spectral fitting of
background-subtracted spectra was performed with XSPEC v12.0
\citep{2004HEAD....8.1629A} and our own shell and TCL scripts to
automate the process as described in \citet{2006A&A...455..903F}.
Because background corrected spectra are not appropriately handled by
C-statistics \citep{2005ApJS..160..319G}, best-fit parameters for the
chosen models were computed by means of the Chi-Squared ($\chi^2$)
minimization. 

We fit our spectra assuming emission by a thermal plasma, in
collisional ionization equilibrium, as modeled by the {\sc APEC} code
\citep{2001ApJ...556L..91S}. Elemental abundances are not easily
constrained with low-statistics spectra and were fixed at Z=0.3
Z$_\odot$, with solar abundance ratios taken from
\citet{1989GeCoA..53..197A}. The choice of sub-solar abundances is
suggested by several  X-ray studies of star forming regions
\citep[e.g.][]{2002ApJ...574..258F,2003A&A...401..543P}.  Absorption
was accounted for using the {\sc WABS} model, parameterized by the
hydrogen column density, N$_{\rm H}$ \citep{1983ApJ...270..119M}. In
Table\,\ref{xspectab} we give best-fit parameters ($\chi_\nu^2$, N$_{\rm H}$,
kT and \Lx) of the sources.

{\small
\begin{table}[!h]
\caption{X-ray spectral fits of \tr sources. The complete version is available 
in the electronic edition of A\&A.}
\label{xspectab}
\begin{tabular}{lllllll}
\multicolumn{7}{l}%
{{\bfseries}} \\
\hline \hline
\multicolumn{1}{l}{N$_{\rm x}$} &
\multicolumn{1}{l}{Cnts.} &
\multicolumn{1}{l}{{Stat.}} &
\multicolumn{1}{l}{log(N$_{\rm H}$)} &
\multicolumn{1}{l}{kT} &
\multicolumn{1}{l}{log(L$_{\rm x}$)} &
\multicolumn{1}{l}{{flag}} \\
\multicolumn{1}{l}{\#} &
\multicolumn{1}{l}{(ph)} &
\multicolumn{1}{l}{($\chi^2_\nu$)} &
\multicolumn{1}{l}{(cm$^{-2}$)}&
\multicolumn{1}{l}{(keV)}&
\multicolumn{1}{l}{(erg/s)} &
\multicolumn{1}{l}{} \\
\hline
   1 &     30 &   $--$ &    $-----$ &   $-----$ & 30.51 &       no-fit  \\
   2 &     36 &   $--$ &    $-----$ &   $-----$ & 30.58 &       no-fit  \\
   3 &    208 &  0.66 &  22.01$\pm$0.16 &  2.07$\pm$0.56 & 31.58 &	 fitted  \\
   4 &     55 &  0.94 &  21.90$\pm$0.31 &  2.96$\pm$2.61 & 30.50 &	 fitted  \\
   5 &     28 &   $--$ &    $-----$ &   $-----$ & 30.48 &       no-fit  \\
   6 &     41 &   $--$ &    $-----$ &   $-----$ & 30.64 &       no-fit  \\
   7 &     47 &   $--$ &    $-----$ &   $-----$ & 31.29 &{\tiny Tr16-19}\\
   8 &     54 &   $--$ &    $-----$ &   $-----$ & 30.76 &       no-fit  \\
   9 &     45 &   $--$ &    $-----$ &   $-----$ & 30.68 &       no-fit  \\
  10 &     25 &   $--$ &    $-----$ &   $-----$ & 30.42 &       no-fit  \\
  11 &     41 &   $--$ &    $-----$ &   $-----$ & 30.72 &       no-fit  \\
  12 &    132 &  1.50 &  21.23$\pm$0.62 &  2.65$\pm$0.94 & 31.08 &	 fitted  \\
  13 &     89 &  0.40 &  21.64$\pm$0.14 &  0.67$\pm$0.93 & 30.83 &	 fitted  \\
  14 &     15 &   $--$ &    $-----$ &   $-----$ & 30.23 &       no-fit  \\
  15 &     41 &   $--$ &    $-----$ &   $-----$ & 30.65 &       no-fit  \\
  16 &     23 &   $--$ &    $-----$ &   $-----$ & 30.40 &       no-fit  \\
  17 &     28 &   $--$ &    $-----$ &   $-----$ & 30.48 &       no-fit  \\
  18 &     31 &  0.80 &  22.32$\pm$0.32 &  1.36$\pm$0.88 & 31.04 &       fitted  \\
  19 &      7 &   $--$ &    $-----$ &   $-----$ & 29.93 &       no-fit  \\
  20 &     30 &   $--$ &    $-----$ &   $-----$ & 30.51 &       no-fit  \\
  21 &     25 &  1.21 &  21.97$\pm$0.65 &  0.29$\pm$0.18 & 30.02 &	 fitted  \\
  22 &    176 &  0.94 &  21.81$\pm$0.22 &  3.32$\pm$1.45 & 31.46 &	 fitted  \\
  23 &     25 &   $--$ &    $-----$ &   $-----$ & 30.43 &       no-fit  \\
  24 &     22 &   $--$ &    $-----$ &   $-----$ & 30.38 &       no-fit  \\
  25 &    177 &  1.87 &  22.00$\pm$0.18 &  2.17$\pm$0.74 & 31.52 &	 fitted  \\
  26 &     44 &  0.90 &  21.70$\pm$0.27 &  1.49$\pm$0.54 & 30.38 &	 fitted  \\
  27 &     51 &  0.50 &  22.06$\pm$0.32 &  2.11$\pm$1.27 & 31.03 &	 fitted  \\
  28 &     77 &  0.47 &  21.28$\pm$0.72 &  1.49$\pm$0.44 & 30.83 &	 fitted  \\
  29 &     36 &   $--$ &    $-----$ &   $-----$ & 30.59 &       no-fit  \\
  30 &     12 &   $--$ &    $-----$ &   $-----$ & 30.12 &       no-fit  \\
\hline
\end{tabular}

\smallskip {Notes: last column flag: Sources with no spectral
information ({\sc no-fit}) have their X-ray luminosities computed by
using an average count-rate to \Lx conversion factor (see
\S\,\ref{lx}). {\sc "Hard tail"} flag refers to the need of more
components in the spectral models. Spectral fit parameters for OB-type
stars are not listed here as they are presented in Table\,\ref{ob}.}
\end{table}}

Except for 28 massive O- and early B-type stars, we fit source spectra
with one-temperature (1T) plasma models using an automated procedure.
In order to reduce the risk of finding a local minimum in the $\chi^2$
spaces, our procedure chooses the best fit among several obtained
starting from a grid of initial values of the model parameters:
log(N$_{\rm H}$)\,=\,21.0, 21.7, 22.0, 22.4, 22.7 and 23.0 cm$^{-2}$
and kT\,=\,0.5, 0.75, 1.0, 2.0, 5.0 keV. best-fit values of
log($N_{\rm H}$)\,$<$\,20.3 cm$^{-2}$ were truncated at 20.3 for two
cases (sources\,\#150 and \#944) because, in the 0.5-8.0 keV energy
range, ACIS spectra are insensitive to lower column densities. In a
similar way, above 10 keV \chandra is not able to discriminate between
such high temperatures. Therefore, 74 best-fit values of kT above 8
keV were truncated to that value. They are indicated with a flag {\sc
hard-tail} in Table\,\ref{xspectab}.

\begin{figure*}[!ht]
\includegraphics[width=8.5cm,angle=0]{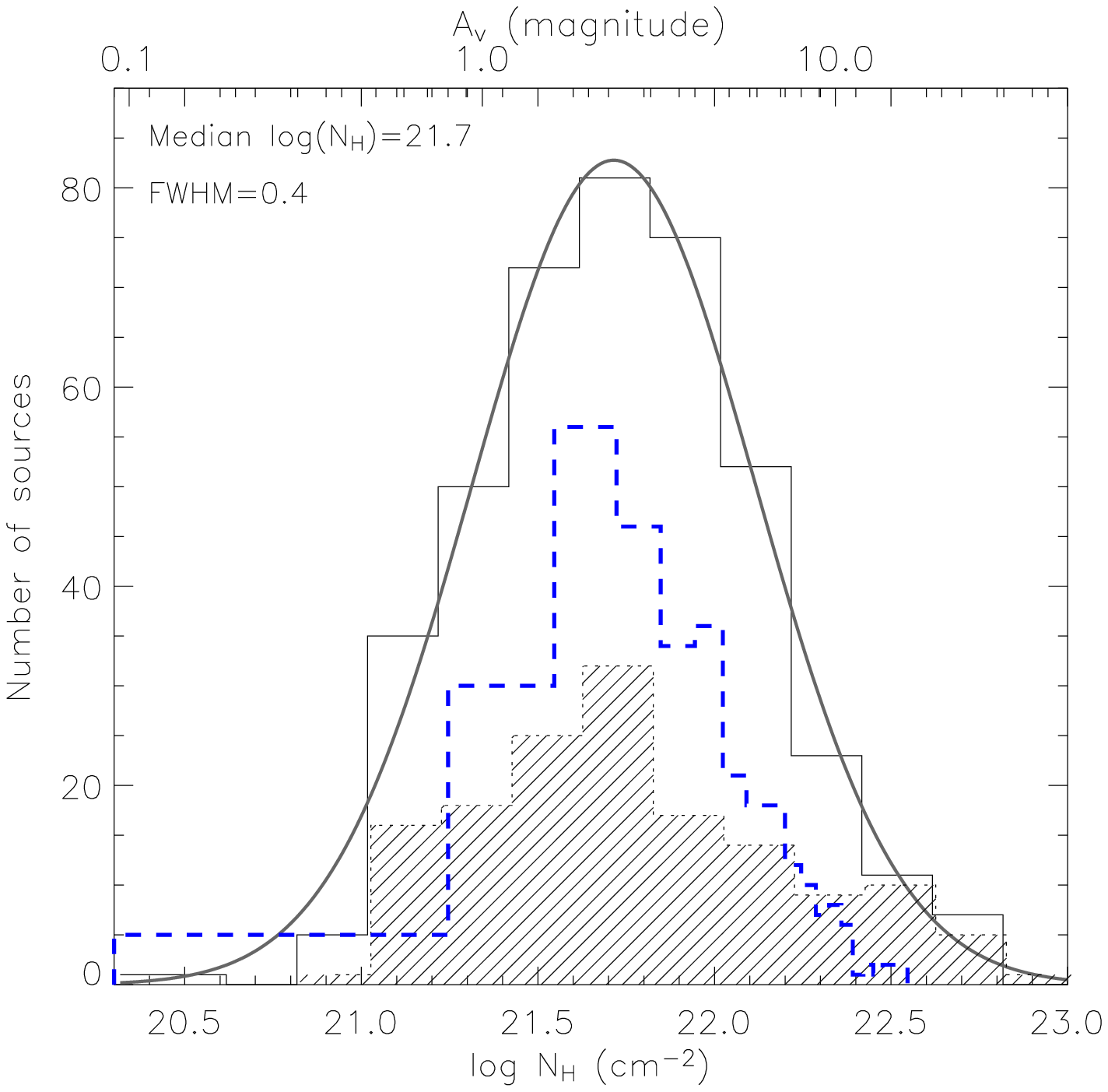}
\includegraphics[width=8.5cm,angle=0]{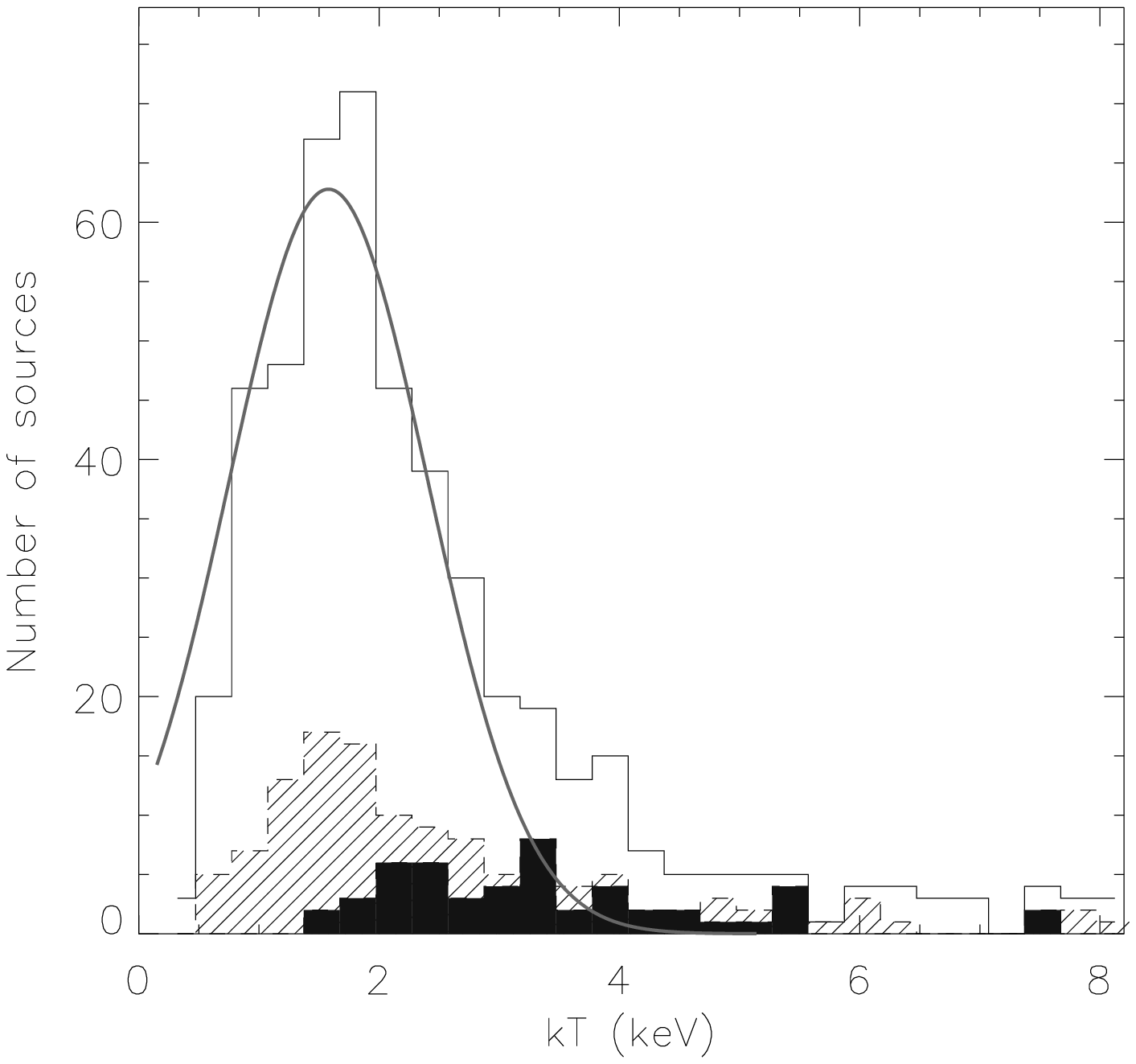}
\caption{{\it Left:} The solid histogram shows the N$_{\rm H}$
(A$_v^{\rm xspec}$) distribution of 415 spectral fitted sources with
2MASS counterparts. The upper scale was converted using an N$_{\rm
H}$/\Av ratio of 1.6$\times$10$^{21}$.  The thick dashed histogram
shows the A$_v^{\rm near-IR}$ mag. distribution for all near-IR
counterparts as was computed in \S\,\ref{near-IR}. Both distributions
peak at A$_v^{\rm xspec}$=3.35 (1$\sigma$=0.4) mag and A$_v^{\rm
near-IR}$=3.26 mag. (1$\sigma$=0.3 dex). The shadowed histogram shows
the N$_{\rm H}$ distribution of 148 sources with no near-IR
counterpart.  {\it Right:} Same as the {\it left} panel for plasma
temperaturess (kT) with, in addition, the distribution for "flaring"
sources (black-filled histogram). The peak of the overal distribution
is at $\sim$ 1.6 keV.}
\label{xspec-nh}
\end{figure*}

Fig.\,\ref{xspec-nh} shows the distribution of best-fit log(N$_{\rm
H}$) values for the 563 fitted sources. They appear to be normally
distributed with a median log\,$N_{\rm H}$\,$\sim$\,21.73 (N$_{\rm
H}$=5.37$\times$10$^{21}$ cm$^{-2}$) and a FWHM of $\sim$0.4 dex. The
log-normal distribution of the sources is indicated with the Gaussian
curve\footnote{Seven sources (\#150, \#305, \#342, \#388, \#692, \#868 and
\#944) appear to have N$_{\rm H}$ below 10$^{21}$ cm$^{-2}$, and are
likely foreground stars.}. The computed median of N$_{\rm H}$
($\sim$5.37$\times$10$^{21}$ cm$^{-2}$) is converted to a median
\Av=3.35 by use of the  \cite{2003A&A...408..581V} relation: N$_{\rm
H}$/\Av\,=\,1.6$\times$10$^{21}$ atoms\,cm$^{-2}$\,mag$^{-1}$. We also
test the relation \Av=0.56\,N$_{\rm H}$+0.23 [N$_{\rm H}$ in
10$^{21}$] \citep{1995A&A...293..889P}, for which the median \Av is
3.23 mag. Both of these values are in good agreement with the median
\Av=3.6 mag computed from our near-IR analysis (see \S\,4.3). The
1$\sigma$ dispersion of the N$_{\rm H}$  distribution is 0.4 dex. It
is translated into typical \Av range between 1.3 to 6.7 mag. of visual
extinction. X-ray sources without near-IR counterparts seem to be
distributed towards higher absorption values (median log(N$_{\rm
H}$)$\sim$21.9 cm$^{-2}$) with respect to those with near-IR
counterparts. Unfortunately, they generally have poor X-ray photon
statistics and consequently a less reliable estimation of their X-ray
spectral parameters.

Unlike the N$_{\rm H}$ distribution, the kT distribution of plasma
temperatures is not log-normal. It peaks at $\sim$1.6 keV, has a
median $\sim$1.95 keV and shows an extended hard tail attributed both to
variable sources (log\,P$_{\rm KS}<-3$) with harder spectra
(median kT=3.25\,keV, as expected from coronal heating processes
involved in flare-like activity) and to highly absorbed sources (i.e.
\Av$\geq$6 mag) showing a median kT$\sim$2.6 keV (while those in the
range 1$\leq$\Av$\leq$6 mag being softer and distributed with a median
kT=1.75 keV). 

Finally, suspected {\it single} massive stars show typically soft
spectra with median kT=0.62 keV, while for known massive binaries
this value rises to kT=2.1 keV, no doubt due to hard X-ray
photons being produced in the colliding wind region (CWR) of the
massive O+OB binaries. Details of X-ray spectral characteristics of
massive stars in the region are presented in  \S\,\ref{obstars}.

\section{X-ray luminosity of stars} \label{lx} 

Unabsorbed X-ray luminosities were computed for those sources with
available spectral fits, for the [0.5-8.0] keV energy range. For
sources with no available and/or reliable spectral fit (119+353 out of
1035) \Lx were computed using a single count-rate to L$_{\rm X}$
conversion factor (CF)\footnote{It was computed as the median ratio
between the individual unabsorbed X-ray luminosities (from best-fit
spectral models) and the source count-rates.}. CF in \tr is
8.3$^{10.3}_{5.2}$$\times$10$^{33}$ ergs/ph. Upper and lower values
correspond to $\pm$1$\sigma$ uncertainties, respectively.

\begin{figure}[!ht]
\centering \includegraphics[width=8.6cm,angle=0]{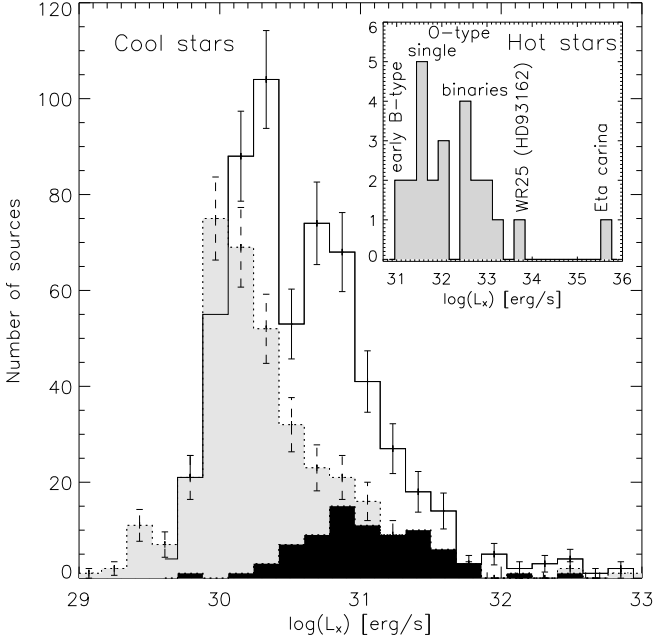}
\caption{The X-ray luminosity distribution for 592 low-mass stars
computed from a single \Lx\,-Count rate conversion factor is shown as
a {\it solid line}. The {\it grey filled} histogram shows the \Lx
distribution for 354 X-ray sources without near-IR counterpart. The 77
variable sources are shown with the {\it black-filled} histogram.
Upper-right inset (Hot stars) shows the unabsorbed \Lx from X-ray
spectral fits (see \S\,\ref{obstars}). Error bars are 1$\sigma$
Poisson errors.} \label{lxhist} 
\end{figure}

In Figure\,\ref{lxhist} we show the X-ray luminosity distribution for
low-mass\footnote{We defined the low mass range M$<$7\,M$_\odot$,
based on the presence of a significant convection envelope (at the
adopted age) that explains X-ray emission from magnetic activity.}
stars, while known OB stars are plotted separately in the upper inset
histogram. The \Lx distribution of sources has been plotted separately
for: $i-$ 592 X-ray sources with near-IR counterpart, not including 28
OB stars and variable sources (thin solid histogram), with a median
log(\Lx) $\sim$30.5 \ergs. The peak of the distribution marks indeed
the completeness limit of our X-ray observation; $ii-$ 354
unidentified sources, except for variable ones, which appear
systematically less luminous than those with 2MASS counterpart, with
median log(\Lx) $\sim$30.2 \ergs\,; $iii-$ 77 variable sources,
showing a median log(\Lx)$\sim$31.1 \ergs, i.e. about 4 times higher
than the observed \Lx for similar stars in a quiescent (non-flaring)
phase. The upper inset in Fig\,\ref{lxhist} shows the unabsorbed \Lx
for 28 massive stars (including 13 binary systems) in the region. By
following results and discussion presented in \S\,\ref{obstars},
binaries appears with typical \Lx over 10$^{+32}$ \ergs, higher than
observed for those suspected single B-type stars. As we will discuss
in \S\,\ref{obstars}, massive binaries have typical X-ray luminosities
\Lx $\geq 10^{+32}$ \ergs, higher than those of single OB stars.

We examine how X-ray activity depends on stellar mass for \tr\,
low-mass stars, and compare the results with those already known for
the ONC and Cyg\,OB2 SFRs. In order to increase statistics, we use \Lx
values computed by means of the average CF. While \Lx and mass for
Cyg\,OB2 stars were computed following the same procedures used here
\citep{2007A&A...464..211A}, for ONC stars they were computed
differently in the literature\footnote{In the ONC, stellar masses were
computed from optical spectra, available for many stars, while \Lx was
obtained for many stars from an X-ray spectral analysis based on a
relatively high photon statistics \citep{2005ApJS..160..319G}.}. For
the sake of homogeneity, we have re-computed \Lx and masses of ONC
stars by using a single count-rate to \Lx conversion factor (CF$_{\rm
onc}$=7.52$\times$10$^{+32}$ erg/ph) and 2MASS photometry,
respectively.

\begin{figure}[!ht]
\centering \includegraphics[width=8.7cm,angle=0]{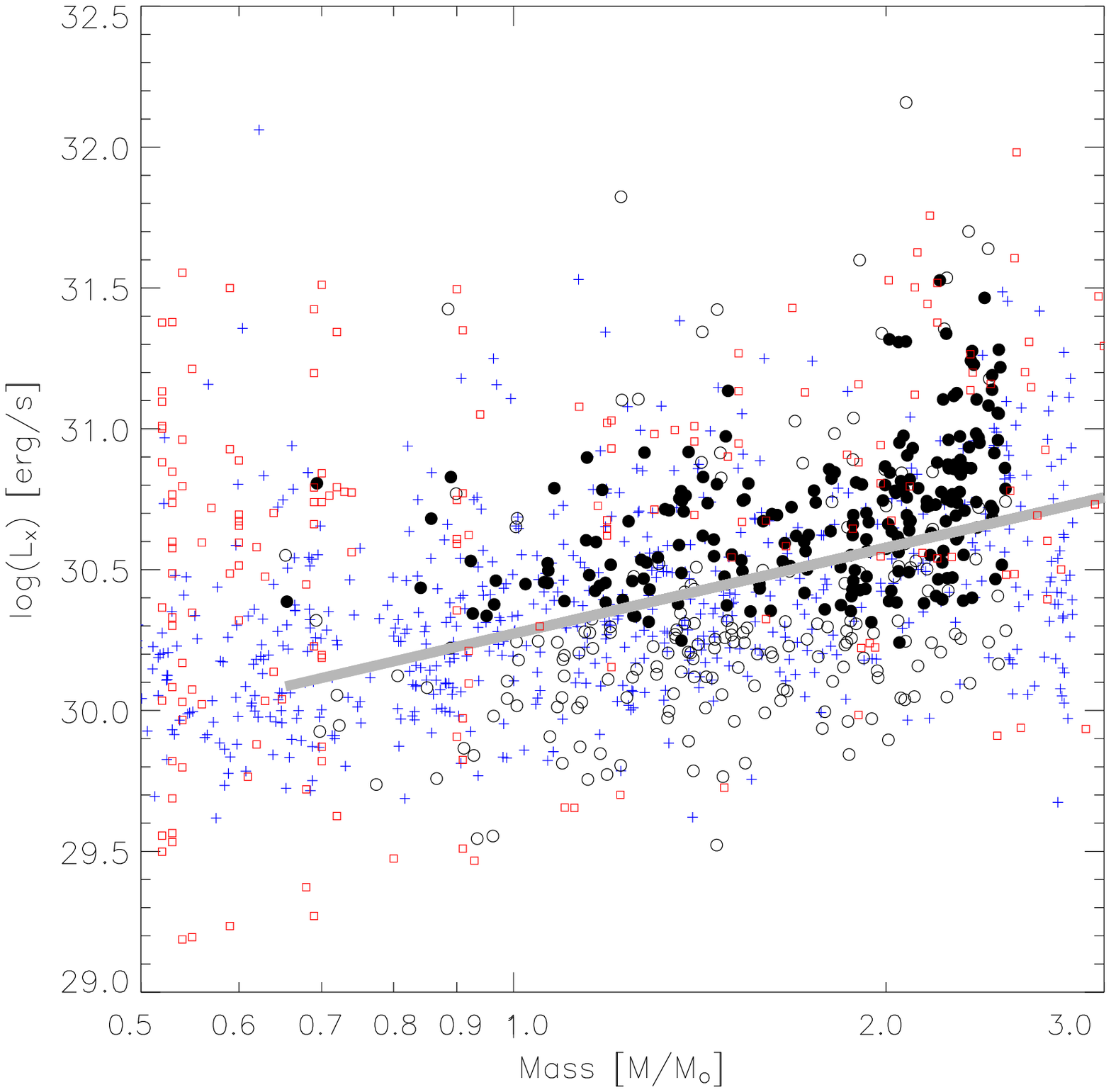}
\caption{X-ray luminosity vs. stellar mass for stars in the near-IR
sample with masses determined from the PMS models of
\cite{2000A&A...358..593S}. Tr16 stars are plotted by open and filled
circles, corresponding to \Lx from CF and spectral fits, respectively.
Crosses indicate Cyg\,OB2 sources \citep{2007A&A...464..211A}, while
small boxes are ONC sources \citep{2005ApJS..160..401P}. The thick
gray line shows the linear regression fit to the Tr16 low-mass
(0.7-2.5 M$_\odot$) stars.} \label{lxscat} 
\end{figure}

\begin{figure}[!ht]
\centering \includegraphics[width=8.7cm,angle=0]{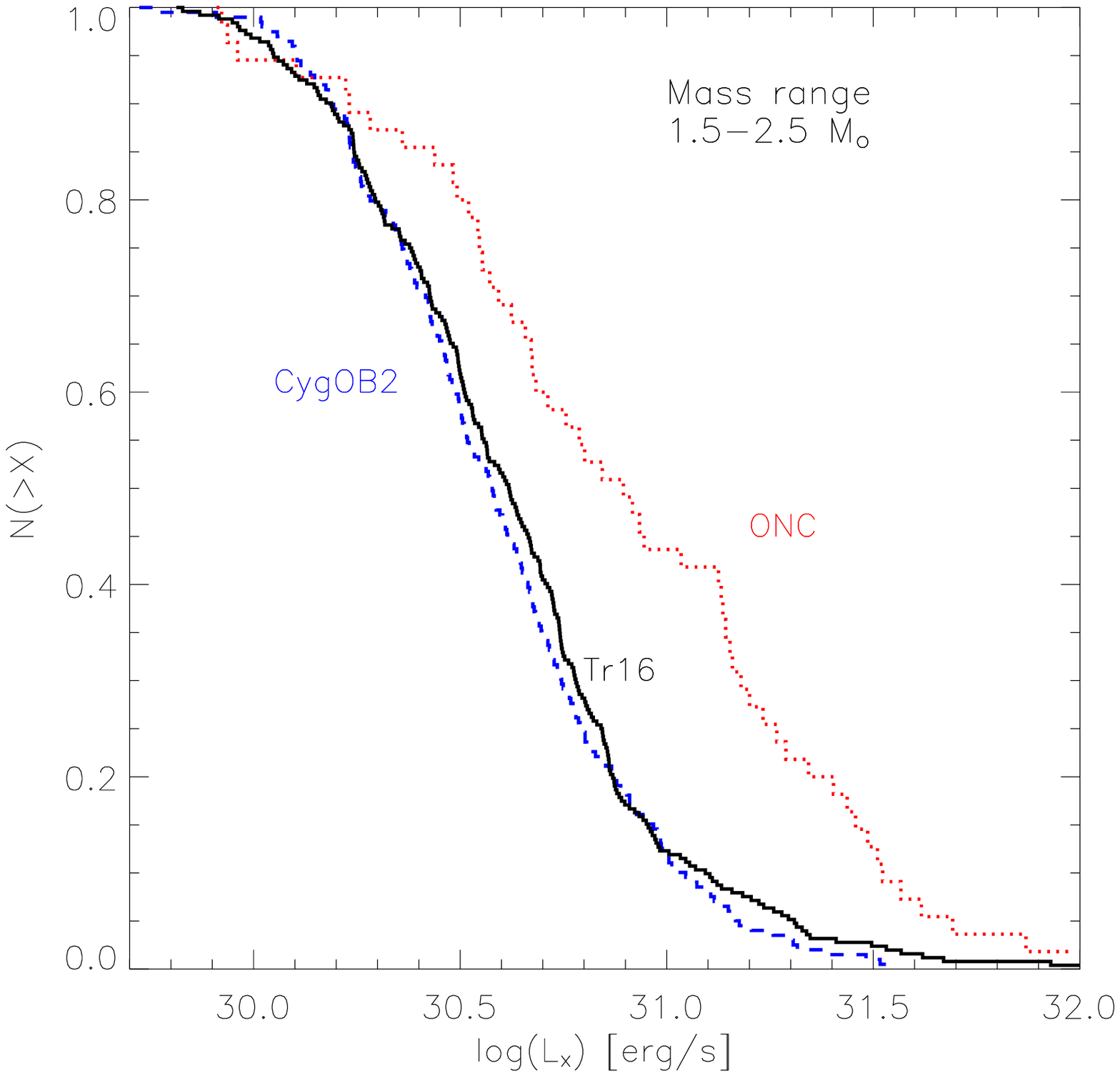}
\caption{Cumulative distributions of X-ray luminosities for \tr
low-mass stars with masses 1.5$\leq\,M/M_\odot\leq\,$2.5 (thick solid
line). Dotted and dashed lines represent the \Lx cumulative
distributions for ONC and CygOB2, respectively, in the same mass
range.} \label{lxcum} 
\end{figure}

In Figure\, \ref{lxscat} we show a plot of \Lx vs. star mass. We use
circles for all 510 Tr16 sources with estimated masses. Filled (329)
and open (181) circles indicate \Lx values computed from spectral
fits and using the CF, respectively. We perform a linear regression
for all sources in the 0.7-2.5 M$_\odot$ range:
log(\Lx)=30.26($\pm$0.11)+1.0($\pm$0.09)log(M/M$_\odot$) with a
standard deviation in the residuals of 0.38 dex. The power-law slope
we find here is in agreement with that found by
\cite{2007A&A...464..211A} for the Cyg~OB2 region:
log(\Lx)=30.33($\pm$0.16)+0.71($\pm$0.13)log(M/M$_\odot$) for masses
in the 0.5-3.0 M$_\odot$ range. This slope also agrees with that we
find for the ONC stars in the mass range 0.5-3.0 M$_\odot$, namely
0.82$\pm$0.09.

Changes in the X-ray activity of stars with different ages of the SFRs
has been previously reported by \cite{2005ApJS..160..390P}. To address
this issue, we computed \Lx detection limits for low-mass stars in the
Tr16\,(3\,Myrs), CygOB2\,(2\,Myrs) and ONC\,(1\,Myrs) observations, as
log(\Lx)$\sim$30.5, 30.3 and 28.5 \ergs, respectively, above which a
source is detectable anywhere in the FOV. A mass-dependent
completeness fraction (f$_{\rm comp}$) for our survey of \tr stars was
then computed by adopting the X-ray Luminosity Functions given by 
\cite{2005ApJS..160..390P}: f$_{\rm comp}$ is $\sim$5\% for 0.1-0.5
M$_\odot$, $\sim$40\% for 0.5-0.9 M$_\odot$ and $\sim$55\% in the
0.9-1.2 M$_\odot$ range. Our survey of Tr\,16 is statistically
complete for masses $\geq$\,1.5\,M$_\odot$. In Figure\,\ref{lxcum} we
present \Lx cumulative distributions for \tr, Cyg\,OB2 and ONC stars
in the mass range 1.5-2.5 M$_\odot$ where all three data sets are
complete. While \tr and Cyg~OB2 show very similar distributions, the
ONC looks quite different with respect to Tr16. The distance between
two distributions (D=0.31) was computed using a two-sample
Kolmogorov-Smirnov test \citep{1992nrfa.book.....P}. We are able to
confirm, with a probability $\geq$99.9\% (log(P$_{\rm ks}$)=-4.15),
that young ONC\,(1\,Myr) stars, with masses in the range 1.5-2.5
M$_\odot$, are intrinsically more luminous in X-rays than their
Tr16\,(3\,Myr) counterparts.

\section{X-rays from massive stars}
\label{obstars}

{\small
\begin{sidewaystable*}
\caption{Stellar parameters and X-ray spectral results for massive
stars.} 
\label{ob} 
\begin{tabular}{lllrllllllllll}
\multicolumn{14}{l}%
{{\bfseries}}\\
\hline \hline
\multicolumn{1}{l}{N$_{\rm x}$} &
\multicolumn{1}{l}{Name} & 
\multicolumn{4}{c}{Stellar parameters} &
\multicolumn{1}{l}{} &
\multicolumn{6}{c}{X-ray spectral parameters} &
\multicolumn{1}{l}{Notes} \\
\cline{3-6} \cline{8-13}
\multicolumn{1}{l}{\#} &
\multicolumn{1}{l}{} &
\multicolumn{1}{l}{Spectral}&
\multicolumn{1}{l}{L$_{\rm bol}$}&
\multicolumn{1}{l}{\mdot} &
\multicolumn{1}{l}{L$_{\rm w}$} &
\multicolumn{1}{l}{} &
\multicolumn{1}{l}{$\chi_\nu^2$/d.o.f} &
\multicolumn{1}{l}{N$_{\rm H}$} &
\multicolumn{1}{l}{Abund.} &
\multicolumn{1}{l}{kT$_1$} &
\multicolumn{1}{l}{kT$_2$} &
\multicolumn{1}{l}{L$_{\rm X}$} &
\multicolumn{1}{l}{} \\
\multicolumn{1}{l}{} &
\multicolumn{1}{l}{} &
\multicolumn{1}{l}{Type}&
\multicolumn{1}{l}{[L$_\odot$]}&
\multicolumn{1}{l}{[M$_\odot$/yr]} &
\multicolumn{1}{l}{[L$_\odot$]} &
\multicolumn{1}{l}{} &
\multicolumn{1}{l}{} &
\multicolumn{1}{l}{[10$^{22}$] cm$^{-2}$} &
\multicolumn{1}{l}{[Z$_\odot$]} &
\multicolumn{1}{l}{[keV]} &
\multicolumn{1}{l}{[keV]} &
\multicolumn{1}{l}{[erg/s]} &
\multicolumn{1}{l}{} \\
\hline
   7&	 Tr16-19&     O9.5V    &47863  & 0.150&33.00 &&  1.01/28 &0.55$\pm$0.25&(1.0)&0.63$\pm$0.17&$-----$	 &1.95$_{0.9}^{2.2}$10$^{31}$&\\
  52&	Tr16-124&	B1V    &23442  &  0.08&8.800 &&  0.98/28 &0.75$\pm$0.33&0.18 &0.49$\pm$0.28&2.27$\pm$1.26&4.13$_{1.4}^{4.2}$10$^{31}$&\\       
  74&	 HD93162&     WN6ha    &1659586&  10.5&5140  &&  1.9/240 &0.63$\pm$0.03&0.54 &0.71$\pm$0.01&2.66$\pm$0.06&5.23$_{5.17}^{5.32}$10$^{33}$&fl-6.7 keV\\
  89&	Tr16-244&      O4If    &851138 &  10.0&7060  &&  0.95/54 &1.50$\pm$0.11&0.53 &0.43$\pm$0.06&1.09$\pm$0.31&1.01$_{0.53}^{1.04}$10$^{33}$&em-lines\\
 136&	 Tr16-11&     B1.5V    &19055  &  0.06&4.400 &&  1.02/17 &0.79$\pm$0.23&(1.0)&0.42$\pm$0.24&3.05$\pm$1.48&4.05$_{1.58}^{5.05}$10$^{31}$&variable\\
 207&	 Tr16-10&	B0V    &37154  & 0.120&22.00 &&  0.90/19 &0.35$\pm$0.31&(1.0)&0.59$\pm$0.13&2.85$\pm$1.50&1.88$_{0.96}^{1.93}$10$^{31}$&\\
 228&	HD93204 &      O5V((f))&309029 &  1.30&780.0 &&  1.08/33 &0.21$\pm$0.07&0.42 &0.50$\pm$0.06&$-----$	 &6.15$_{1.55}^{9.96}$10$^{31}$&\\
 242&	HD93205A&O3.5V((f))    &575440 &  2.50& 1870 &&  2.12/74 &0.66$\pm$0.04&1.2  &0.26$\pm$0.02&1.73$\pm$0.60&1.89$_{1.06}^{2.26}$10$^{33}$&(A),primary,CW\\
 242&	HD93205B&	O8V    &91201  & 0.260&91.00 &&   "    &   "	  &"	   &"	    &"        &"	&(B),secondary,CW\\
 281&	 Tr16-21&	O8V    &91201  & 0.260&91.00 &&  1.09/14 &0.66$\pm$0.31&(1.9)&0.33$\pm$0.18&$-----$	&2.92$_{0.09}^{3.41}$10$^{31}$&\\
 286&	 Tr16-14&     B0.5V    &29512  & 0.100&17.00 &&  $-----$ &$-----$    &$-----$&$-----$	   &$-----$	&$-----$ &not fitted\\
 352& CPD-592600&  O6V((f))    &208929 & 0.800&410.0 &&  1.07/45 &0.54$\pm$0.12&1.0  &0.29$\pm$0.05&0.78$\pm$0.11&3.12$_{1.76}^{3.70}$10$^{32}$&\\
 407&CPD-592603A&  O7V((f))    &138038 & 0.400&170.0 &&  1.08/23 &0.58$\pm$0.14&(1.0)&0.31$\pm$0.07&$-----$	&1.29$_{0.87}^{1.79}$10$^{32}$&(A),primary\\
 407&CPD-592603B&     O9.5V    &47863  & 0.150&33.00 &&   "    &   "	  &"	   &"	    &"        &"	&(B),secondary\\
 407&CPD-592603C&     B0.2V    &37153  & 0.120&22.00 &&   "    &   "	  &"	   &"	    &"        &"	&(C),open\\
 489&	  Tr16-5&	B1V    &23442  &  0.08&8.800 &&  1.13/18 &0.23$\pm$0.09&(1.0)&2.87$\pm$0.88&$-----$	 &1.26$_{1.08}^{1.43}$10$^{31}$&variable\\
 649& \etacar-MS&   O2If       &3715354&  8.32& 6780 &&  3.75/342&7.39$\pm$0.39&4.25 &0.27$\pm$0.32&7.16$\pm$3.6&4.02$_{0.6}^{8.5}$10$^{35}$&(A),fl-6.7 keV\\
 649& \etacar-B &    O5V       &309029 &  1.30&780.0 &&   "    &   "	  &"	   &"	    &"        &"	&(B),CW,variable\\
 687&	 Tr16-23&	O7V    &138038 & 0.400&170.0 &&  0.78/23 &0.60$\pm$0.09&(1.0)&0.46$\pm$0.07&$-----$	&5.53$_{4.34}^{6.51}$10$^{31}$\\
 688&	  Tr16-9&     O9.5V    &47863  & 0.150&33.00 &&  1.02/13 &0.48$\pm$0.29&(1.0)&0.30$\pm$0.16&1.0$\pm$0.23&3.78$_{1.40}^{4.20}$10$^{31}$\\
 689&  HDE303308&  O4V((f))    &467735 &  2.00&1410  &&  1.80/49 &0.54$\pm$0.05&(1.0)&0.25$\pm$0.03&0.52$\pm$0.09&6.16$_{6.10}^{7.87}$10$^{32}$&em-lines\\
 707&	  Tr16-3&     O8.5V    &72443  & 0.220&67.00 &&  0.74/16 &0.25$\pm$0.15&(1.0)&0.33$\pm$0.15&$-----$	&1.10$_{0.19}^{1.27}$10$^{31}$&\\
 729&CPD-592628A&     O9.5V    &47862  & 0.150&33.00 &&  1.08/17 &0.93$\pm$0.79&(1.0)&0.11$\pm$0.17&0.96$\pm$0.74&3.11$_{1.79}^{6.05}$10$^{32}$&(A) primary\\
 729&CPD-592628B&     B0.3V    &37153  & 0.120&22.00 &&   "    &   "	  &"	   &"	    &"        &"	&(B) secondary\\
 730&	 Tr16-22&     O8.5V    &72443  & 0.220&67.00 &&  1.3/64  &1.07$\pm$0.10&(1.9)&0.35$\pm$0.07&1.86$\pm$0.12&7.71$_{7.06}^{9.34}$10$^{32}$&prob. binary\\
 759&	 Tr16-74&	B1V    &23442  &0.0800&8.800 &&  $-----$ &$-----$    &$-----$&$-----$	   &$-----$	&$-----$ &not fitted\\
 803&	HD93343 &	 O7V(n)&138038 & 0.400&170.0 &&  0.93/32 &0.78$\pm$0.09&(1.0)&0.28$\pm$0.03&$>$4.0	&2.96$_{2.63}^{3.92}$10$^{32}$&em-lines\\
 807&	 Tr16-76&	B2V    &15488  &  0.05&2.200 &&  $-----$ &$-----$    &$-----$&$-----$	   &$-----$	&$-----$ &not fitted\\
 808&CPD-592635A&	O8V    &91201  & 0.260&91.00 &&  0.88/28 &0.58$\pm$0.12&(1.0)&0.28$\pm$0.03&0.90$\pm$0.17&1.40$_{1.09}^{1.86}$10$^{32}$&(A),SB2\\
 808&CPD-592635B&     O9.5V    &47862  & 0.150&33.00 &&   "    &   "	  &"	   &"	    &"        &"	&(B) SB2\\
 812&CPD-592636A&	O7V    &138038 & 0.400&170.0 &&  1.05/33 &0.74$\pm$0.09&(1.0)&0.25$\pm$0.04&0.87$\pm$0.12&3.56$_{2.28}^{4.66}$10$^{32}$&(A),(SB2),em-lines\\
 812&CPD-592636B&	O8V    &91201  & 0.260&91.00 &&   "    &   "	  &"	   &"	    &"        &"	&(B),SB2\\
 812&CPD-592636C&	O9V    &58884  & 0.180&47.00 &&   "    &   "	  &"	   &"	    &"        &"	&(C),SB1\\
 854& CPD-592641&	O5V    &309029 &  1.30&780.0 &&  0.87/36 &0.87$\pm$0.08&(1.0)&0.25$\pm$0.07&0.56$\pm$0.11&5.48$_{3.55}^{7.63}$10$^{32}$&em-lines\\
 888&	Tr16-115&     O8.5V    &72443  & 0.220&67.00 &&  1.01/10 &0.54$\pm$0.32&(1.0)&0.28$\pm$0.18&0.54$\pm$0.32&3.06$_{0.87}^{3.34}$10$^{31}$&\\
 997&   FO15A   &  05.5V((f))  &309029 &  1.30&780.0 &&  0.83/29 &1.15$\pm$0.15&(1.0)&0.75$\pm$0.10&$>$7.5	 &1.05$_{0.82}^{1.31}$10$^{32}$& (A),SB2,em-lines\\
 997&   FO15B   &  08.5V       &91201  & 0.260&91.00 &&   "    &   "	  &"	   &"	    &"        &"	&  (B),SB2\\

\hline																		
\end{tabular}
																	
\smallskip {Notes: {it "em-lines"} reefers to stars with emission lines
in their optical spectra. {\it "CW"} notes those binaries with a
colliding wind region. {\it "Open"} reefers to those long period
binaries, i.e. with well-separated components. Components (A), (B),
(C) refer to system components sorted according to mass decrease.}
\end{sidewaystable*}
}

A variety of different physical mechanisms are responsible for the
observed X-ray emission in OB-type stars. The most widely accepted
explanation invokes multiple small-scale shocks in the inner layers of
their radiation-driven stellar winds
\citep[e.g.][]{1997A&A...322..878F}. In recent years it has gained
importance a plasma heating model known as {\it magnetically channeled
wind shock} (MCWS) \citep[][]{2003ApJ...595..365S,
2005AIPC..784..239O, 2005ApJ...634..712G}. Moreover, WR+OB and/or O+OB
{\it interacting wind binary systems} produce an excess of X-ray
emission from the CWR \citep{1992ApJ...386..265S, 2000ApJ...538..808Z,
2002A&A...388L..20P}.

\begin{figure}[!ht]
\includegraphics[width=8.5cm,angle=0]{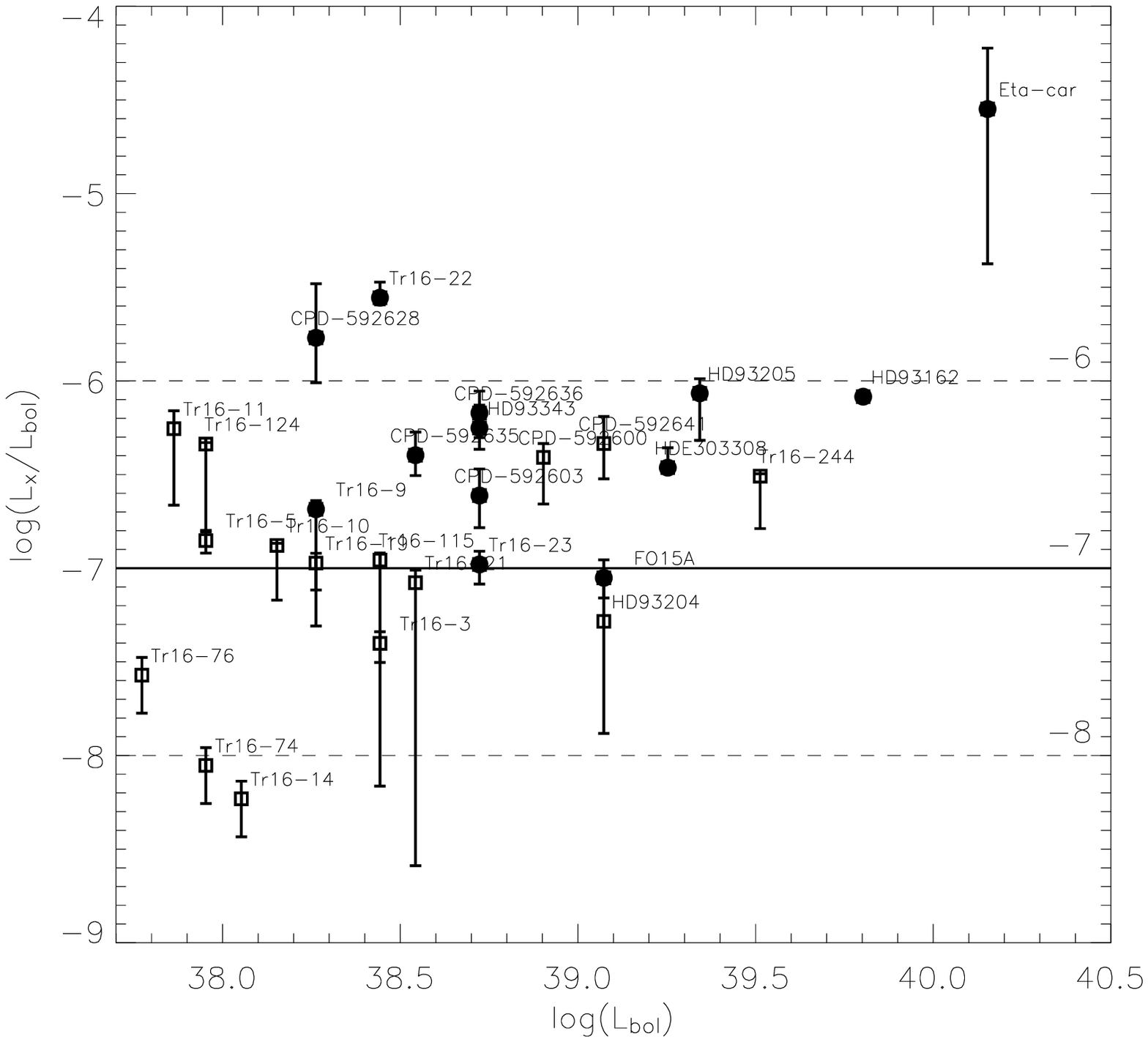}
\caption{log(\Lx/\Lbol) versus log(\Lbol) relation for massive stars
in Tr16. {\it Open boxes} refer to suspected single stars and {\it
filled circles} to known binaries. Notes: $i-$ Because of flare-like
variability (see \S5.1) observed in Tr16-\#5 (B1.5V) and Tr16-\#11
(B1V), their X-ray emission is probably dominated by a unknown
low-mass star companion. $ii-$ The star Tr16-\#22 (O8.5V+?) has the
highest log(\Lx/\Lbol) ratio ($\sim$\,-5.56) among all MS OB-type
stars of the region.}
\label{ob_stars}
\end{figure}

Given the relatively large number of massive OB-type stars in the
region, its appropiate to show in Table\,\ref{ob} most relevants
stellar parameters of the massive stars (\lbol\,, \mdot, and L$_{\rm
wind}$ computed as ${1 \over 2}\mdot$v$_\infty^2$)\footnote{Adopted
values were taken from \cite{2006MNRAS.367..763S}-(table\,1)}. We
computed \Lx by use of an absorbed ({\sc wabs}) thermal plasma model
({\sc apec}). Metal abundance was fixed at Z\,=\,0.3\,Z$_\odot$ in
fitting faint sources ($<$\,100 ph.), while it was left as a free
parameter for the remaining cases (see Table\,\ref{ob}). In
Fig\,\ref{ob_stars} we show the \Lx/\Lbol relation. In spite of the
observed scatter, one sees that the median \Lx/\lbol\, for binaries
(filled circles) is about 7 times larger ($\sim$8.3$\times$10$^{-7}$)
than that of suspected single stars
($\sim$1.1$\times$10$^{-7}$)\footnote{Because of observed flare-like
variability in Tr16-\#5 and \#11, we suspect that most of X-rays
emission comes from a low-mass companion, and therefore we discard
these sources in computing the median \Lx\,.}. Four of the 13 known
binaries in the analysis are well separated systems: the LBV \etacar,
a probable long period binary with P$\sim$2026$\pm$2 days (Daminelli
et\,al., 2007), the Wolf-Rayet star WR25 (HD\,93162), a 208-days
period binary system \citep{2006A&A...460..777G} and  HDE\,303308,
resolved as a binary system with a component separation of about
$\sim$38 AU projected along the fine guidance sensor (FGS)
y-axis\footnote{It was unresolved along the FGS x-axis down to 20 AU,
suggesting an eccentric orbit.}. The stars Tr16\,\#9 (O9.5\,V+?) and
Tr16\,\#23 (O7\,V+?) show photometric variability probably related to
a secondary component \citep{2004AJ....128..323N}, and thus we
consider them also to be binaries.

Besides of the several works \citep[e.g.][]{1979ApJ...234L..55S,
1982ApJ...256..530S, 1995RMxAC...2...97C, 2003MNRAS.346..704C,
2003ApJ...589..509E, 2004ApJ...612.1065E, 2007ApJ...656..462S} about
the origin of the observed X-ray emission on massive stars of this
region, this goal is far out of the scope of our paper and not
extensively discused here.

\section{Summary and conclusion} \label{concl}

We report here results of a deep {\em Chandra} X-ray observation
pointed toward the $\sim$3\,Myr old star forming region \tr. Source
detection was performed using the PWDdetect code, identifying 1035
X-ray sources in the 17'$\times$17' ACIS-I FOV. Most of these seem to
be outside the obscured V-shaped region of dust and gas. Star
formation in this part of the masked region has probably been
disrupted and/or diminished as the stellar winds are blocked inside
the cloud, due to the efficiency of photo-evaporation processes caused
by interactions of the nearby hot massive stars with the dense dust
and gas structures. 

Data extraction was performed using the semi-automated IDL-based {\sc
ACIS Extract} package, which is well suited to the analysis of
observations of crowded fields such as ours.

The X-ray source list was cross-identified with optical and near-IR
(2MASS) catalogs: 28 X-ray sources (of 44 within the FOV) were
identified with optically characterized OB members of \tr and 760 with
2MASS sources. Among these latter sources almost all are believed to
be \tr members. About 90 X-ray sources without optical/NIR
counterparts are estimated to be of extragalactic nature (AGNs), while
the remaining X-ray sources with no counterpart are likely associated
with members that are fainter than the 2MASS completeness limit.

In order to characterize the previously unidentified likely cluster
members with NIR counterparts, we placed them on NIR color-magnitude
(K$_{\rm s}$\,vs.\,H-K$_{\rm s}$) and color-color (H-K$_{\rm
s}$\,vs.\,J-H) diagrams. A first estimate of interstellar extinction
was obtained adopting a 3\,Myr isochrone for the  low- and
intermediate-mass stars and assuming that  O- and early B-type stars
lie on the MS. We find a median visual absorption for OB stars of
\Av$\sim$2.0 mag, while low mass likely members seem to be slightly
more absorbed, \Av$\sim$3.6 mag. We also use the 3\,Myr isochrone and
the J magnitude to estimate masses of likely members assuming that
they share the same distance and absorption. Our sample of X-ray
selected members with near-IR counterparts reaches down to
M=0.5-0.6\,M$_\odot$, and is likely complete down to
$\sim$1.5\,M$_\odot$. From the H-K$_{\rm s}$\,vs.\,J-H diagram we
estimate that $\sim$15\% (51/339) of low-mass stars have NIR excesses,
finding it to be quite a high percentage with respect to the 2 Myr old
Cyg\,OB2. We believe that the disk fraction in young SFRs is more
dependent on spatial morphology of gas and dust around massive stars,
which may enhance photo-evaporation, and thus shorten disk lifetimes,
than on the total number of massive stars in the region.

At least 77 sources, i.e. $\sim$7.4\%, were found to be variable
within our observation with a confidence level greater than 99.9\%.
Only three of the 28 detected O- and early B-type stars were detected
as variable during our 90-ksec observation, in spite of the high
statistics of the OB stars' light curves. These exceptions are the
known binary O3.5V+O8V star HD93205 (our source \#242), showing a
rather linear decay of the count rate during the observation plus a
short-term variability, the B1.5\,V star Tr16-11 (source \#136) and
the B1V star Tr16-5 (source\#489). The latter two show a flare-like
variability probably related to unresolved low-mass companions.

We modeled the ACIS X-ray spectra of sources with more than 20 photons
and where f$_{\rm cont}\,<$1. We assumed an absorbed single-component
thermal emission model. The median log(N$_{\rm H}$) of the sources is
21.73 (cm$^{-2}$). This value agrees well with the median \Av computed
from the near-IR diagram. The median kT of low-mass stars is 2.6 keV.
Sources associated with O- and early B-type stars are instead quite
soft (median kT: 0.60 keV). Absorption corrected X-ray luminosities of
OB stars were calculated from the best-fit spectral models. O and
B-type stars are the most luminous, with L$_{\rm x}$=
2.5$\times10^{30}$-6.3$\times10^{33}$  erg\,s$^{-1}$. Their X-ray and
bolometric luminosities are in rough agreement with the relation 
L$_{\rm x}$/L$_{\rm bol}=10^{-7}$, albeit with an order of magnitude
dispersion. Low mass stars have L$_{\rm x}$ ranging between $10^{30}$
and $10^{31}$ \ergs (median \Lx = $2.8\times10^{30}$). Variable low
mass stars are on average 0.5 dex brighter (log(\Lx)$\sim$31.0 \ergs).
These X-ray luminosities are consistent with those of similar mass
stars in the slightly younger (2\,Myr) Cyg~OB2 region. However, in the
mass range 1.5-2.5 M$_\odot$, the ONC\,(1~Myr) shows higher X-ray
activity level than observed in \tr stars in the same mass range. We
believe that the age-\Lx activity connection is an acceptable
explanation of this result. 

\begin{acknowledgements}

We thank the referee, Marc Gagne, for his time and many useful
comments that improved this work. This publication makes use of data
products from the Two Micron All Sky Survey, which is a joint  project
of the University of Massachusetts and the Infrared Processing and
Analysis  Center/California Institute of Technology, funded by the
National Aeronautics and Space  Administration and the National
Science Foundation. J.F.A.C acknowledges support by the Marie Curie
Fellowship Contract No. MTKD-CT-2004-002769 of the project "{\it The
Influence of Stellar High Energy Radiation on Planetary Atmospheres}",
and the host institution INAF - Osservatorio Astronomico di Palermo
(OAPA). J.F.A.C is a researcher member of the Consejo Nacional de
Investigaciones Cient\'ificas y Tecnol\'ogicas (CONICET)-Argentina and
acknowledges support from this institution. G.\,M., F.\,D. and S.\,S.
acknowledge financial support from the Ministero dell'Universit\`a e
della Ricerca research grants, and ASI/INAF Contract I/023/05/0.

\end{acknowledgements}

\bibliographystyle{astron}
\bibliography{jfac_accepted}

\end{document}